\documentclass[%
 reprint,
 amsmath,amssymb,
 aps,
]{revtex4-2}

\usepackage{graphicx}% Include figure files
\usepackage{dcolumn}% Align table columns on decimal point
\usepackage{bm}% bold math

\usepackage{hyperref}
\usepackage{graphicx}
\usepackage{float}
\usepackage{caption}
\usepackage{subcaption}
\usepackage[shortlabels]{enumitem}
\expandafter\let\csname equation*\endcsname\relax
\expandafter\let\csname endequation*\endcsname\relax
\usepackage{amsmath,amsthm,amssymb,cancel,esint}
\usepackage{mathtools}

\newcommand{\RNum}[1]{\uppercase\expandafter{\romannumeral #1\relax}}

\usepackage{array}
\usepackage{graphicx}
\usepackage{multirow}
\usepackage{color,soul}
\usepackage{epsfig}

\newcommand\MyBox[2]{
  \fbox{\lower0.75cm
    \vbox to 1.7cm{\vfil
      \hbox to 1.7cm{\hfil\parbox{1.4cm}{#1\\#2}\hfil}
      \vfil}%
  }%
}

\usepackage{subcaption}
\usepackage{array,graphicx}
\usepackage{booktabs}
\usepackage{pifont}
\usepackage[table]{xcolor}
\usepackage{tabularx}
\usepackage{adjustbox}

\newcommand*\rot{\rotatebox{90}} 
\newcommand*\rotnine{\rotatebox{90}}

\begin{document}

\preprint{APS/123-QED}

\title[Gravity Spy]{A new method to distinguish gravitational-wave signals from detector noise transients with Gravity Spy}

\author{Seraphim Jarov$^1$}\email{sjarov94@student.ubc.ca} \author{Sarah Thiele$^2$} \author{Siddharth Soni$^3$} \author{Julian Ding$^{1,4}$} \author{Jess McIver$^1$} \author{Raymond Ng$^4$} \author{Rikako Hatoya$^5$} \author{Derek Davis$^6$}
\address{$^1$ Department of Physics and Astronomy, University of British Columbia, Vancouver, British Columbia, V6T1Z4, Canada}
\address{$^2$ Department of Astrophysical Sciences, Princeton University, 4 Ivy Lane, Princeton, NJ 08544, USA}
\address{$^3$ LIGO Lab, Massachusetts Institute of Technology, Cambridge, MA 02139, USA}
\address{$^4$ Department of Computer Science, University of British Columbia, Vancouver, British Columbia, V6T1Z4, Canada}
\address{$^5$ University of California, Los Angeles Electrical and Computer Engineering Dept., Los Angeles, CA 90095}
\address{$^6$ LIGO, California Institute of Technology, Pasadena, CA 91125, USA}

\date{\today}% It is always \today, today,

\begin{abstract}
The Advanced LIGO and Advanced Virgo detectors have enabled the confident detection of dozens of mergers of black holes and neutron stars. However, the presence of detector noise transients (glitches) hinders the search for these gravitational wave (GW) signals. We prototyped a restructuring of Gravity Spy's classification model to distinguish between glitches and astrophysical signals. Our method is able to correctly classify three-quarters of retracted candidate events in O3b as non-astrophysical and 100\% of the confirmed astrophysical events as true signals. This approach will inform candidate event validation efforts in the latest observing run.
\end{abstract}

\maketitle

%\tableofcontents

\section{Introduction}

Since the first observing run (O1) in 2015, the Advanced LIGO and Advanced Virgo detectors have seen significant improvements in detector sensitivity~\cite{LIGO_O2_O3,low_latency}. As a result, the rate of confirmed gravitational wave (GW) detections has increased steadily with each observing run~\cite{aLIGO,Virgo_second_gen,Discover_GWs_w_aLIGO}. However, GW detector data also contains a high rate of non-astrophysical noise transients (glitches). These glitches come in a variety of morphologies~\cite{gwtc3,gwtc21,glitch_paper_for_GW150914,LSC_glitch_paper,DIRECT}. They often arise at different rates between each detector; and are a result of many known environmental and instrumental factors as well as unknown sources~\cite{LIGO_O2_O3,Berger_2018}. As a result, detector data can become heavily polluted, making searches for GWs and accurate extraction of source properties substantially more difficult~\cite{LIGO_O2_O3,O1_CBC}. 
Frequent detector glitches can bias analyses when glitches occur close in time to candidates, or mimic the form of real GW events and produce false positive candidate events~\cite{Glitch_mimic_gw}. 
For example, in the third observing run (O3), just under a third of open public alerts (OPAs) were retracted~\cite{gwtc3,gwtc21,emfollowup}. 
To aid in event validation and potentially reduce the rate of retracted event candidates, we propose a quick and accurate method based on single-detector GW strain data for determining whether a glitch is present in the data that is robust to loud astrophysical events. 
This method will complement existing signal-vs-glitch classification programs, including the use of Q-occupancy to determine data quality~\cite{Soni:2023pxg}, GWSkyNet~\cite{Cabero_2020,Abbott_2022} which requires multiple detectors, iDQ~\cite{iDQ}, which requires auxiliary witness data, and $p_{astro}$~\cite{Kapadia_2020} reported in open public alerts~\cite{gwtc3,gwtc21}. 

A widely-used machine learning image classifier for GW detector characterization, Gravity Spy~\cite{GSpy,BAHAADINI2018172,Glanzer:2022avx}, has achieved high accuracies for classification of detector glitches~\cite{Glitch_mimic_gw,Soni_2021}. The Gravity Spy project leverages a convolutional neural network (CNN) image classifier using time-frequency representations of detector data called qscans~\cite{GSpy,Chatterji}. Qscans are also commonly referred to as omega scans~\cite{LdvW} and spectrograms~\cite{Glitch_mimic_gw,GSpy}.

We restructured Gravity Spy's CNN to provide compact binary coalescence (CBC) GW signal-vs-glitch classifications which will allow for rapid rejection of false candidate events using GW strain data. We emphasize that this method does not require auxiliary witnesses or multiple detectors. 
Our approach adds new capability relative to current methods, such as iDQ~\cite{iDQ}, which requires auxiliary witness data, or GWSkyNet~\cite{Cabero_2020,Abbott_2022}, which requires data from multiple detectors to classify a candidate event. The purpose of the original glitch-classification Gravity Spy CNN model~\cite{GSpy} was to classify detector glitches that arise in the LIGO data stream, which it accomplishes with high accuracy. However, the classifier was not trained to distinguish between glitches and astrophysical GW signals.
In particular, when we first tested our new model architecture with the original Gravity Spy training set, we found that simulated GW signals that come from both high mass ($>50M_\odot$ total mass) and low mass ($<50M_\odot$ total mass) mergers tend to be misclassified as glitches that appear similar in appearance, duration, and frequency range~\cite{thiele2021}. 
Examples of these types of glitches and simulated signals are presented in Figures \ref{fig:low_mass_chirp} and \ref{fig:high_mass_chirp}. 
This is particularly problematic for CBC sources with total mass below $30M_\odot$, as long duration ($>0.5$ seconds) glitches such as light scattering that mimic these signals have been common in GW detector data during previous observing runs~\cite{Virgo_second_gen,gwtc3,Soni_2021,LIGO:2020zwl,Glanzer:2023hzf, O2_O3_DC}. We therefore aimed to expand Gravity Spy beyond its current capabilities by improving signal-vs-glitch classification on top of its glitch classification framework.
We also tackled the challenge of future-proofing our classification network by considering sources with masses higher than previous detections~\cite{HM_CBC_mot}, as the expected sensitivity improvements~\cite{LIGO_O2_O3,low_latency} in future observing runs could result in signals from a broader mass spectrum being detected. 

To match the accuracy of Gravity Spy's glitch classification accuracy for the signal-vs-glitch case, we introduced a new structure to Gravity Spy's CNN to automate the process described in \cite{Davis_2020}. 
Motivated by commonalities in glitch types that are known to mimic the appearance of GW signals from certain mass ranges, we split the Gravity Spy CNN into two classifiers, one focusing on distinguishing long duration (i.e. low mass) CBC signals from glitches with similar time-frequency morphology, and one focusing on short duration (i.e. high mass) CBC signals and similar glitches. 
We trained the low mass classifier on simulated GW signals with total mass between $3 M_\odot$ and $50 M_\odot$ 
and the high mass classifier on simulated GW signals with total mass between $50 M_\odot$ and $250 M_\odot$. 
We trained the low mass and high mass classifiers to be robust against shifts in time-frequency image centering by augmenting training set images with time offets. 
Other qscan processing techniques have also been explored resulting in a third classifier constructed to classify signals in the $250M_\odot$ to $350M_\odot$ mass range. 
We present a proof-of-principle GW candidate event validation tool that performs reliably and efficiently. 

\section{Methods}\label{sec_2}

\subsection{Generating Simulated GWs}\label{sec_2.1}

We first identified a $64$ second segment of quiet LIGO Livingston detector data\footnote{We note that this short data segment is sufficiently representative of nominal data for this proof-of-principle study.} to act as baseline noise we can inject a simulated waveform into using the PyCBC package~\cite{PyCBC}. 
For the purpose of our study, we simulated signals drawing parameters from a total mass range of $3M_\odot$ to $350M_\odot$, a signal-to-noise-ratio (SNR) range of $3$ to $35$, and a range of $-0.95$ to $0.95$ for both component spins to span the parameter space of most likely GW sources for the expected O4 observing run. 
We injected each generated waveform into the quiet timeseries data, then used Gravity Spy to process the data as a series of four qscans, as described in~\cite{GSpy}.
This workflow can supply qscans of GWs with a variety of parameters for mass, spin, and SNR. We used this to test Gravity Spy and our prototype image classifiers leveraging time-frequency representations of GW detector data.

\subsection{Generating new training sets}

In our study, we constructed and trained two main classifiers which focus on low-mass and high-mass CBC signals and the types of glitches that have similar time-frequency morphology in a qscan~\cite{Glitch_mimic_gw,GSpy,Soni_2021,LdvW,cabero_blips}, as outlined in Table \ref{tab:glitches}. 
To determine which glitch classes have similar time-frequency morphology, we first ran the original Gravity Spy on a range of simulated GW signals, as described in Section~\ref{Test_GSpy}, and included glitch classes where the original Gravity Spy model confused simulated signals in each mass range for these classes.

\begin{table*}[t]
\centering
\begin{tabular}{|c|c|}
\hline
\textbf{Low mass classifier classes} & \textbf{High mass classifier classes} \\ \hline
\rowcolor[HTML]{D9D9D9} 
Blip & Blip \\ \hline
Low frequency blip & Low frequency blip \\ \hline
\rowcolor[HTML]{D9D9D9} 
Scratchy & Koi Fish \\ \hline
No Glitch & Tomte \\ \hline
\rowcolor[HTML]{D9D9D9} 
GW ($3-50\ M_{\odot}$ total mass CBC) & GW ($50-300\ M_{\odot}$ total mass CBC) \\ \hline
\end{tabular}
\caption{The signal and glitch types we used to retrain the Gravity Spy CNN for optimal signal versus glitch classification, for each classifier.}\label{tab:glitches}
\end{table*}

Equipped with our GW simulation workflow, we retrained Gravity Spy's CNN (we use the same CNN as presented in \cite{Zevin_2017}) on enriched training sets that increased representation of simulated GW signals outside the mass and SNR ranges present in the original training set. 
Our enriched training set contains an equal representation of each class, with roughly 750 examples per classifier, as equal representation of classes been shown to increase the robustness of a training set~\cite{hastie_ml}. 
We discuss the classes included in our method in the next section.

\section{Challenging types of glitches for different CBC mass ranges}\label{sec_3}

Here we consider particularly challenging cases for a signal-vs-glitch classifier intaking qscan time series data, especially cases where the time-frequency morphology of glitches is similar to low mass or high mass CBCs. 

\subsection{Low mass CBCs}
\begin{figure*}[ht]
\centering
\begin{tabular}{ccc}
\begin{tabular}[b]{@{}c@{}}
\includegraphics[width=0.3\textwidth]{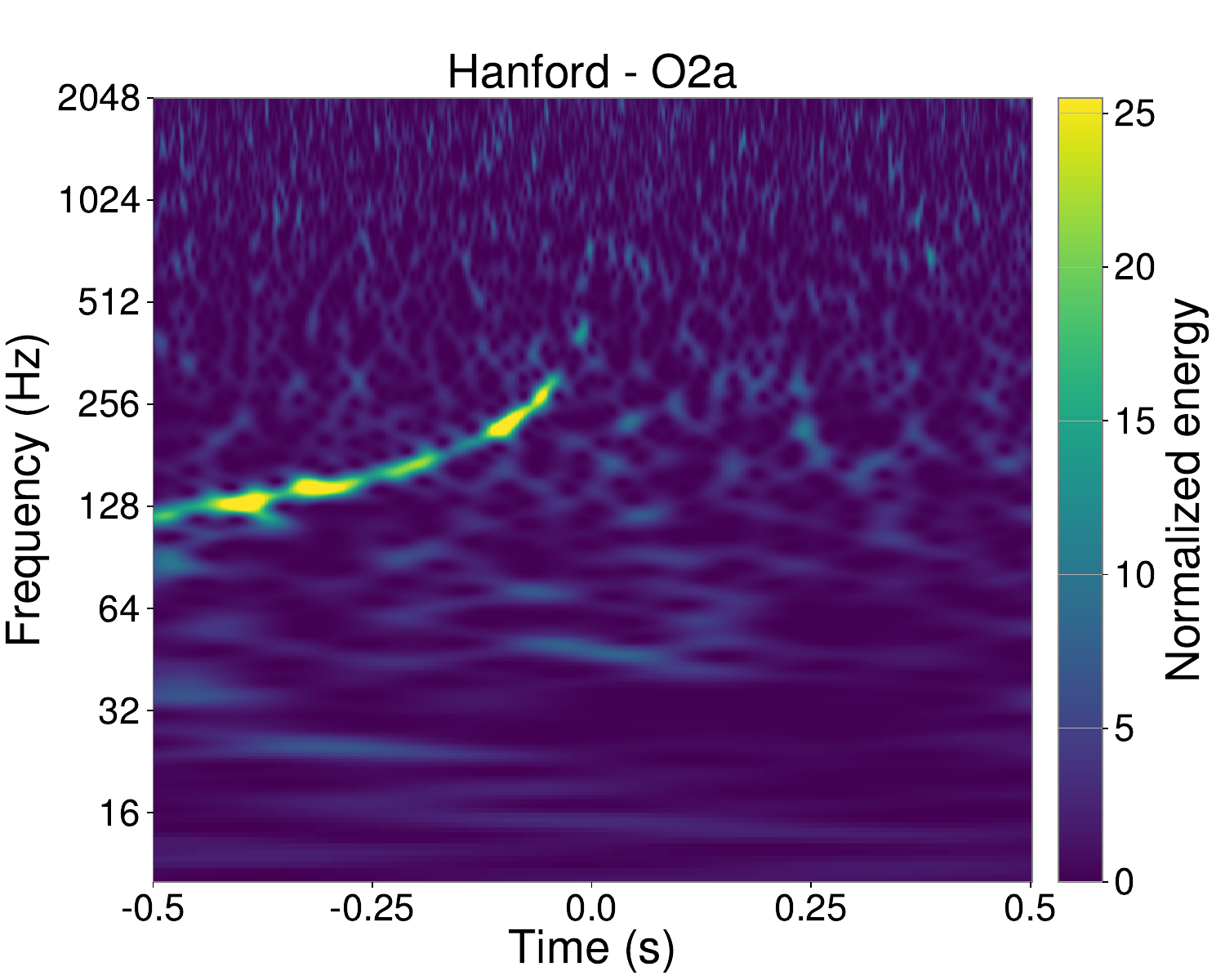} \\
\fontsize{12}{14}\selectfont\textbf{I}
\end{tabular} &
\begin{tabular}[b]{@{}c@{}}
\includegraphics[width=0.3\textwidth]{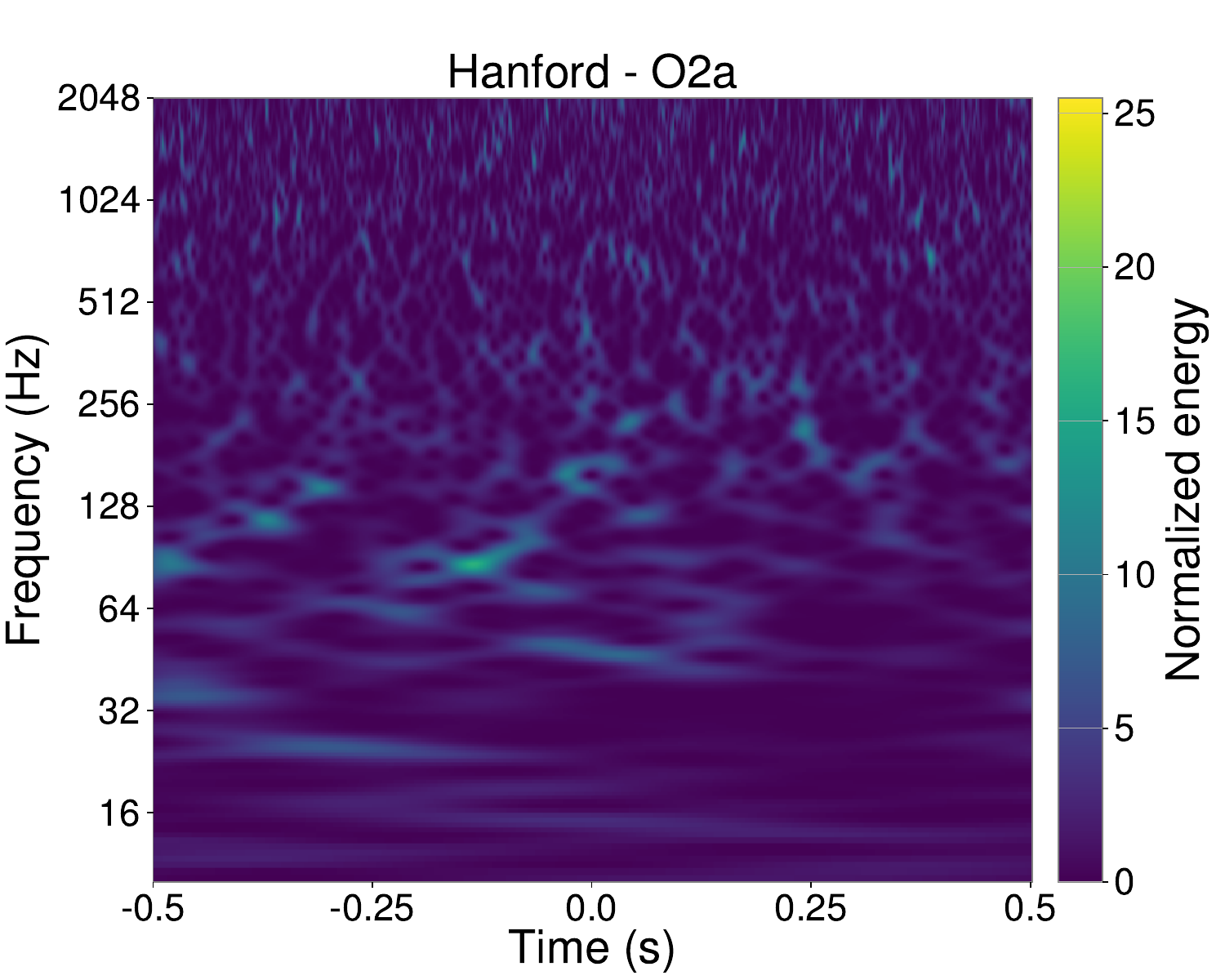} \\
\fontsize{12}{14}\selectfont\textbf{II}
\end{tabular} &
\begin{tabular}[b]{@{}c@{}}
\includegraphics[width=0.3\textwidth]{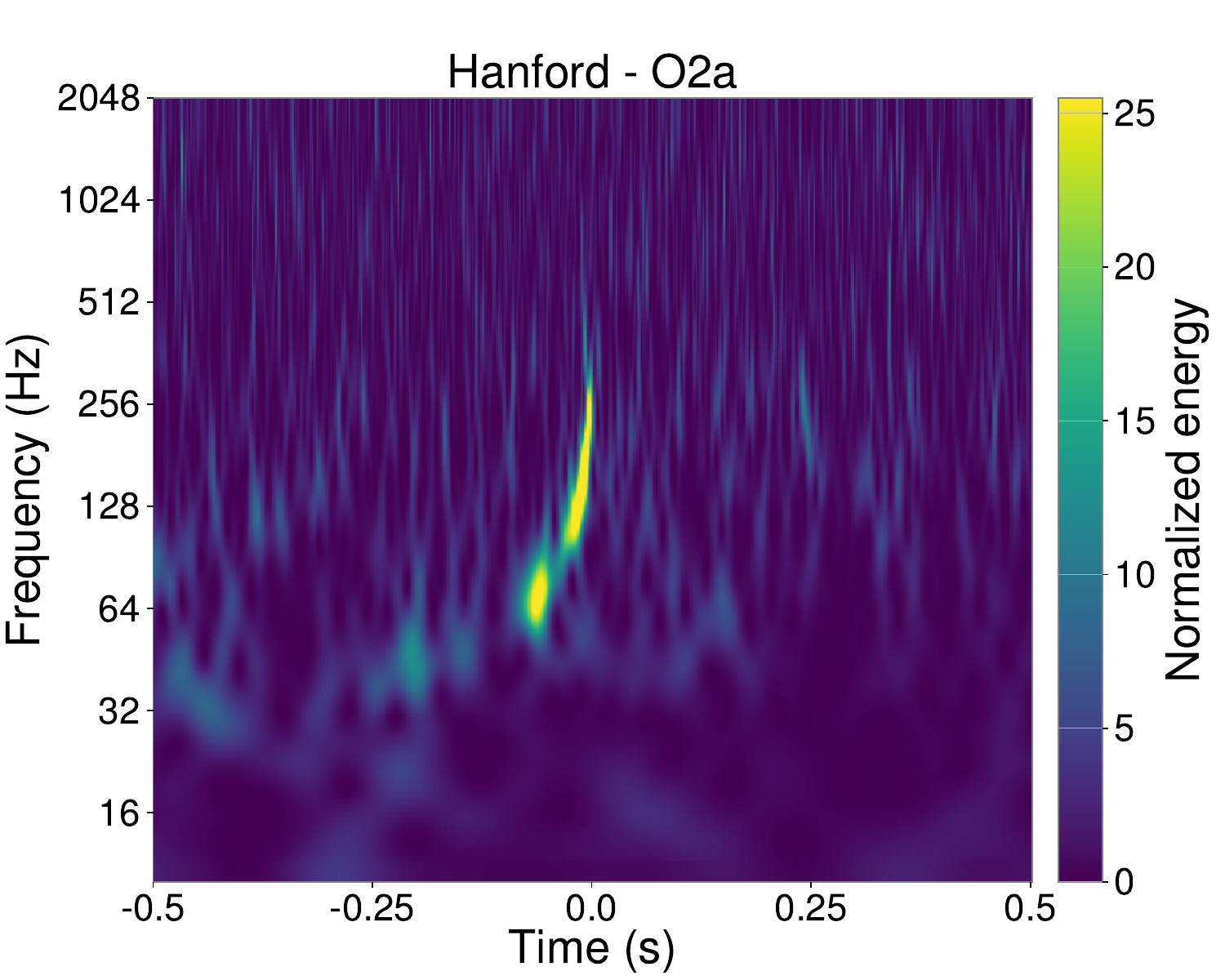} \\
\fontsize{12}{14}\selectfont\textbf{III}
\end{tabular} \\
\begin{tabular}[t]{@{}c@{}}
\includegraphics[width=0.3\textwidth]{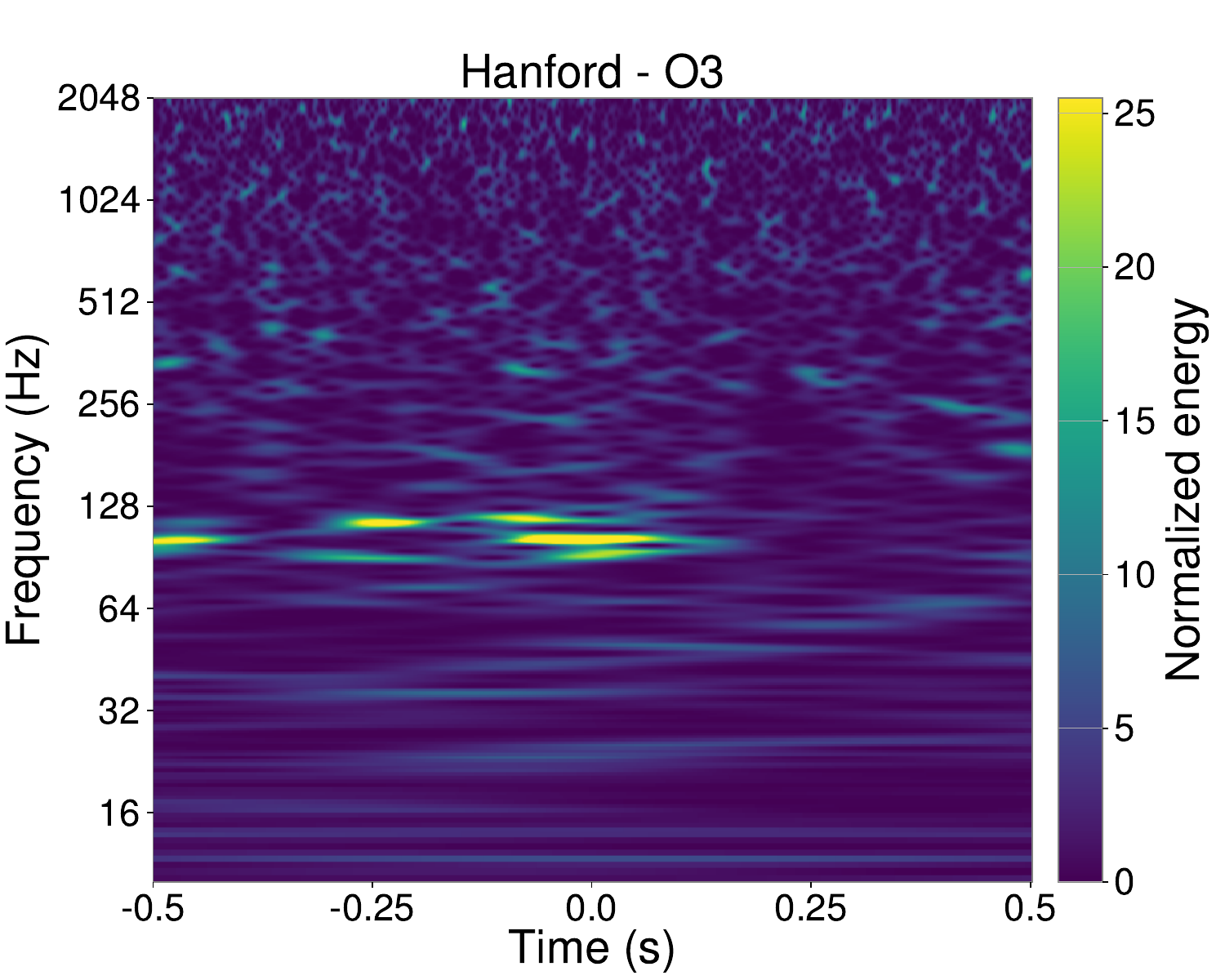} \\
\fontsize{12}{14}\selectfont\textbf{A}
\end{tabular} &
\begin{tabular}[t]{@{}c@{}}
\includegraphics[width=0.3\textwidth]{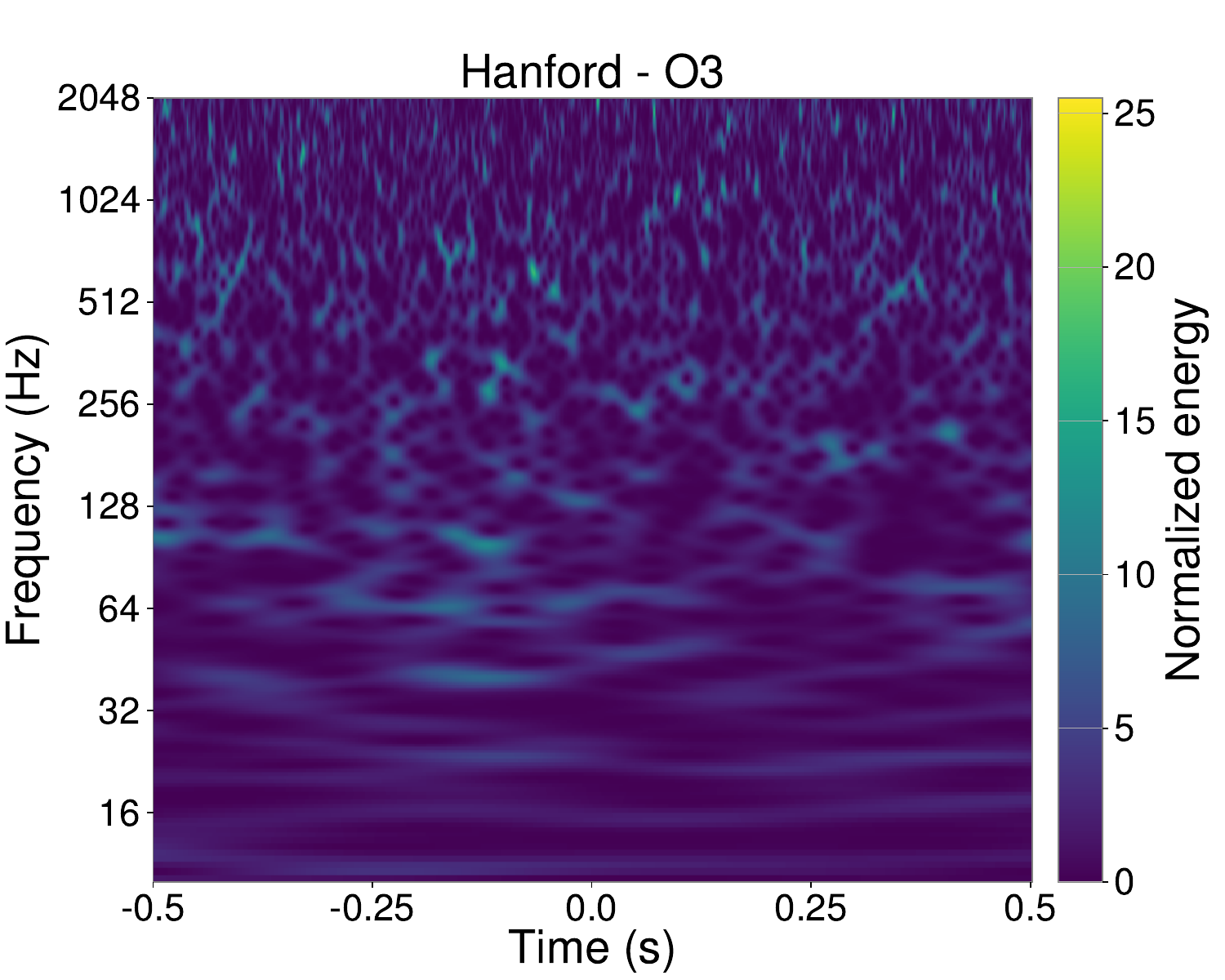} \\
\fontsize{12}{14}\selectfont\textbf{B}
\end{tabular} &
\begin{tabular}[t]{@{}c@{}}
\includegraphics[width=0.3\textwidth]{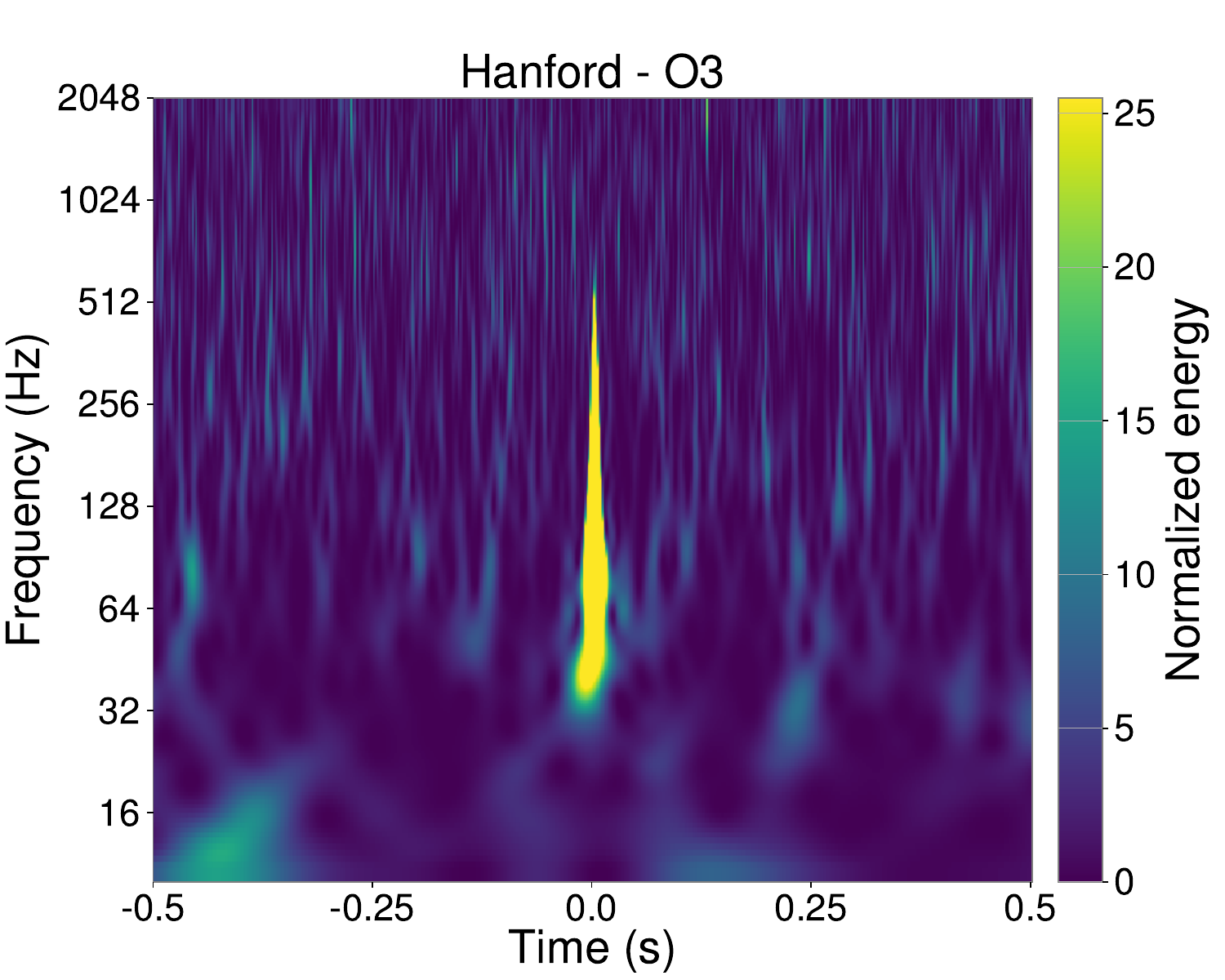} \\
\fontsize{12}{14}\selectfont\textbf{C}
\end{tabular}
\end{tabular}
\caption{Example qscans of low mass CBC GW signals and glitches confused by the original Gravity Spy model. 
A $5 M_{\odot}$ total mass GW source (\textbf{\RNum{1}}) is often misclassified as a scratchy glitch (\textbf{A}), which has a similar frequency range and is also relatively long in duration. 
A GW source of total mass $16 M_{\odot}$ (\textbf{\RNum{2}}) is often misclassified as `no glitch' (\textbf{B}). 
This is expected behavior; there is little difference in a qscan between a low SNR GW signal and Gaussian detector noise.  
Lastly, a GW source of $46.5 M_{\odot}$  (\textbf{\RNum{3}}) is commonly confused for a blip glitch (\textbf{C}) due to similar time-frequency morphology in a qscan.}
\label{fig:low_mass_chirp}
\end{figure*}

CBC sources with a total mass below $50M_{\odot}$ are longer in duration and generally manifest as lower in energy in a qscan relative to higher mass (short duration) signals. We present examples in Figures \ref{fig:low_mass_chirp} \textbf{\RNum{1}}, \textbf{\RNum{2}}, and \textbf{\RNum{3}}.

The original Gravity Spy training set contained very few low mass GW examples. 
As a result, the original Gravity Spy model tends to confuse simulated low mass CBCs with glitches that are also long in duration and have similar frequency content. 
The original Gravity Spy model also tends to classify low mass CBCs, especially simulated GWs with low SNR, as `no glitch'. An example is shown in Figure \ref{fig:low_mass_chirp} (\textbf{\RNum{2}} and \textbf{B}). 
However, we note that this is not an undesirable outcome. 
A `no glitch' classification is equivalent to a GW classification for the purposes of event validation; neither class is actionable for further data quality investigation. 

\subsection{High mass CBCs}

\begin{figure*}[ht]
\centering
\begin{tabular}{ccc}
\begin{tabular}[b]{@{}c@{}}
\includegraphics[width=0.3\textwidth]{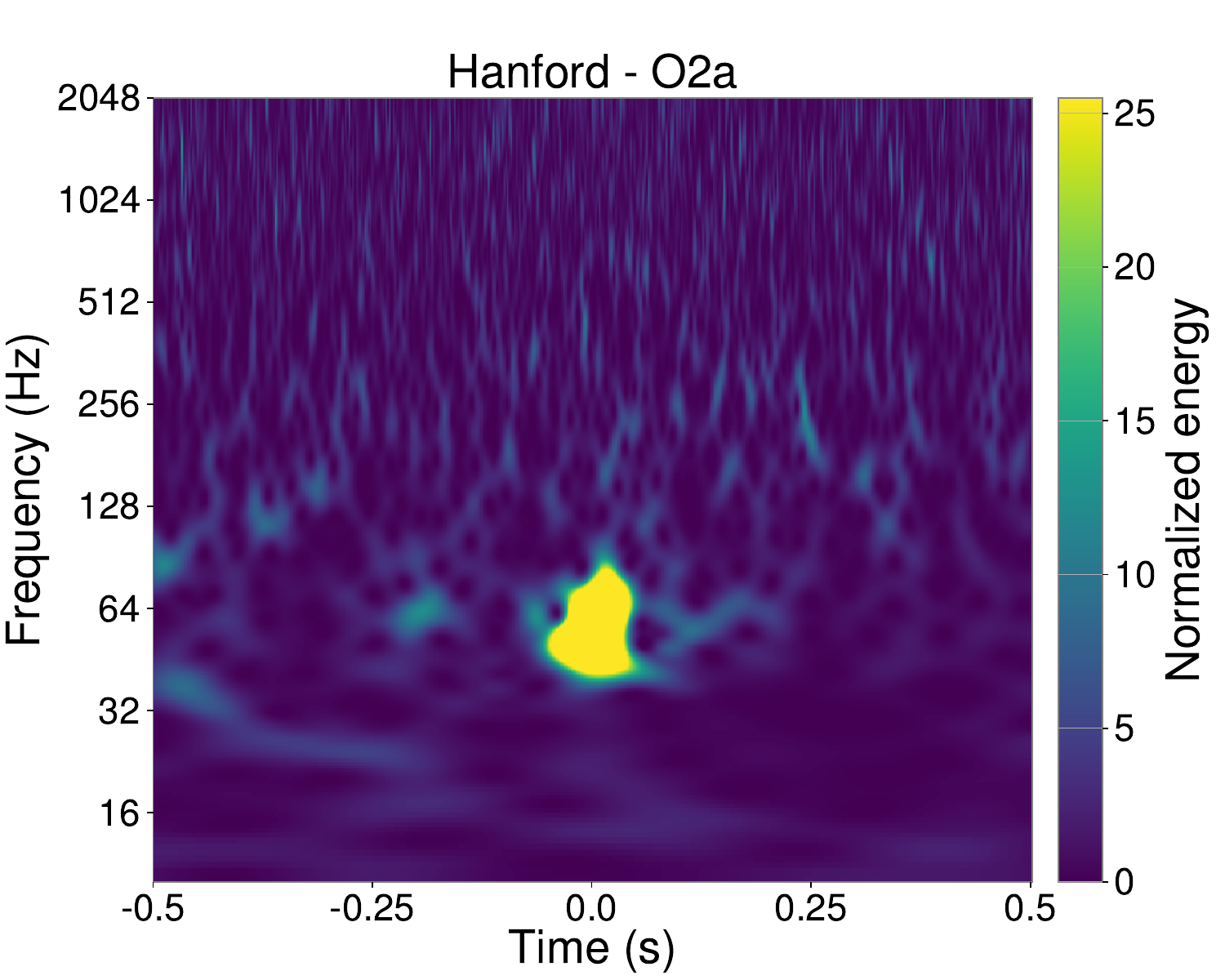} \\
\fontsize{12}{14}\selectfont\textbf{I}
\end{tabular} &
\begin{tabular}[b]{@{}c@{}}
\includegraphics[width=0.3\textwidth]{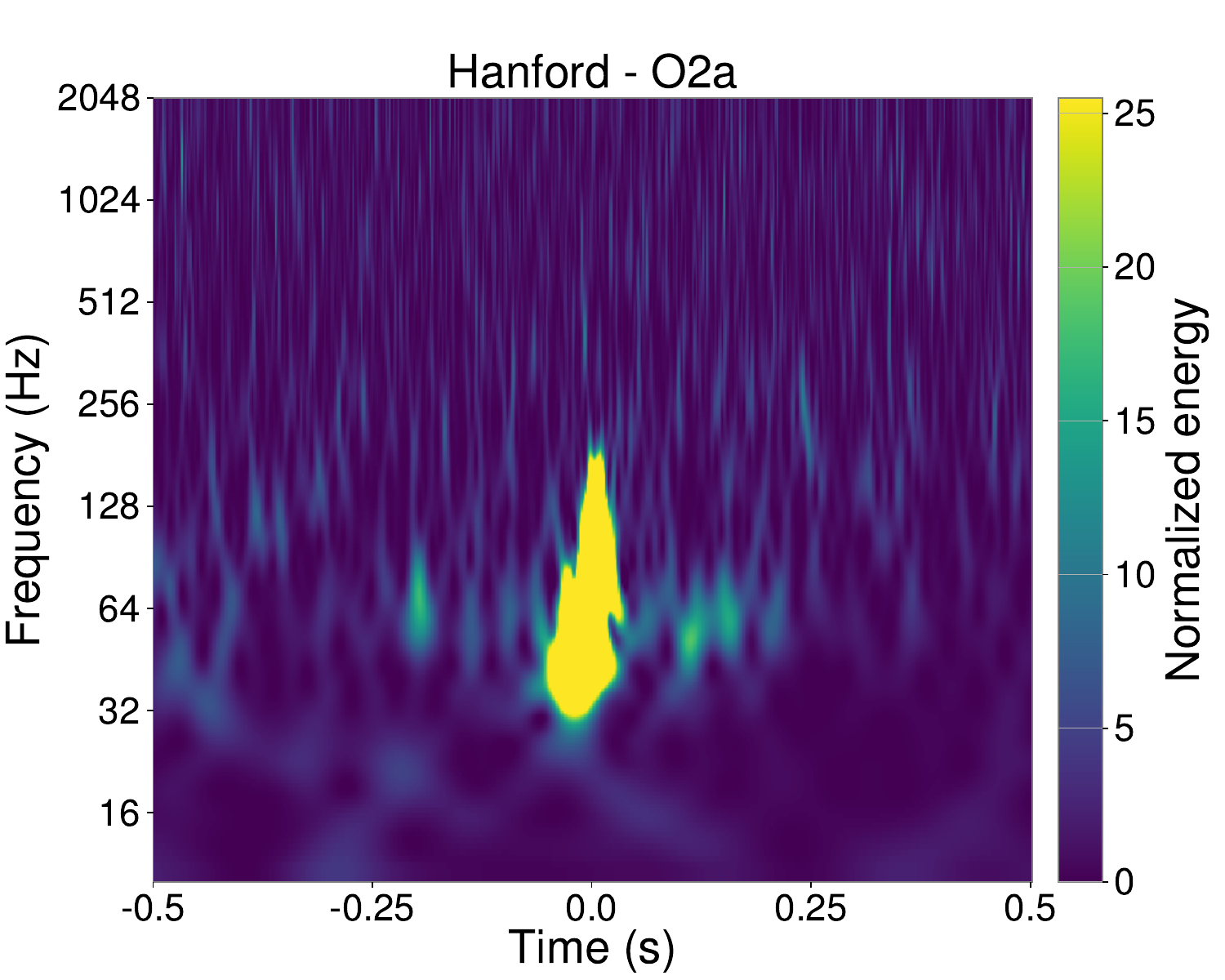} \\
\fontsize{12}{14}\selectfont\textbf{II}
\end{tabular} &
\begin{tabular}[b]{@{}c@{}}
\includegraphics[width=0.3\textwidth]{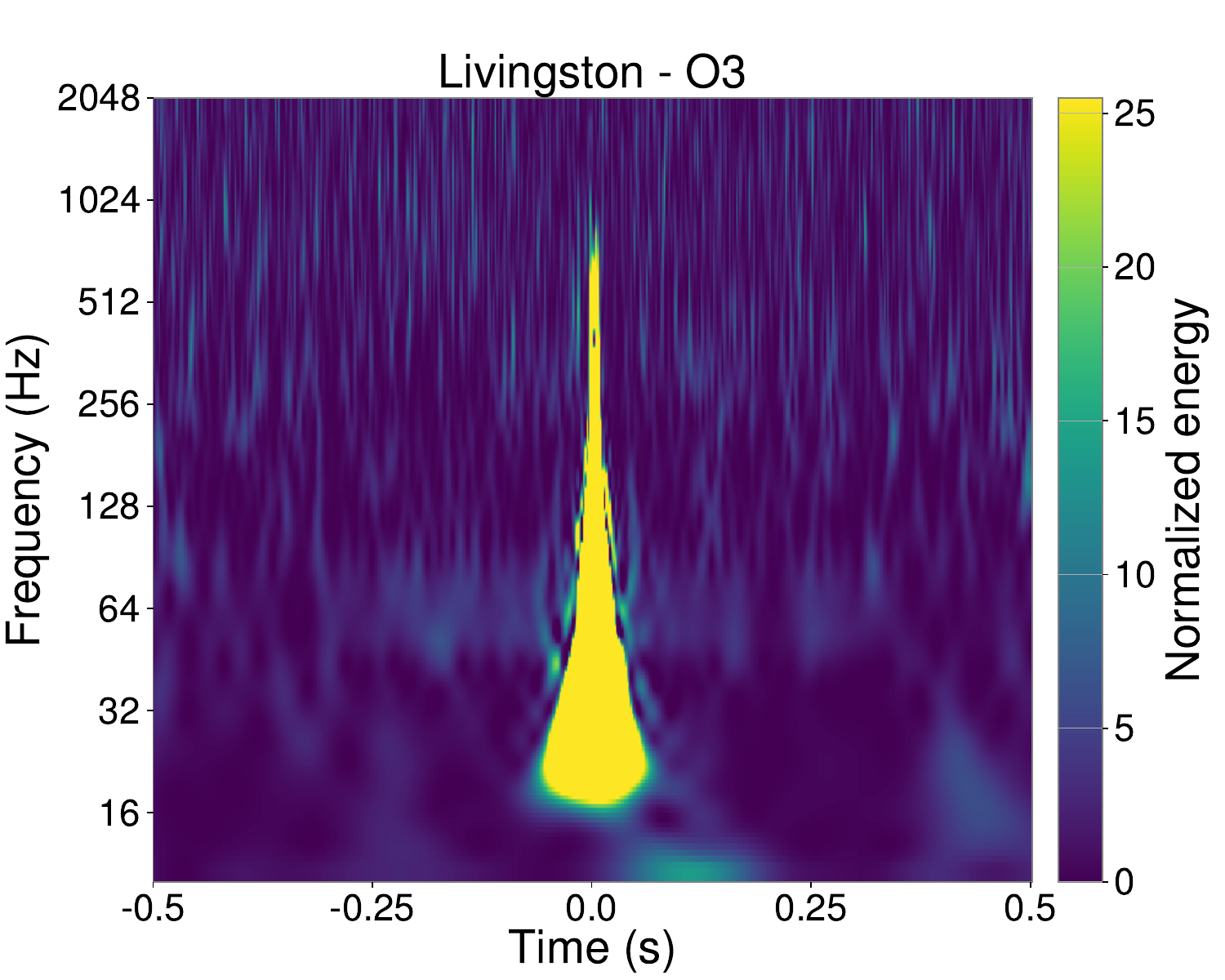} \\
\fontsize{12}{14}\selectfont\textbf{D}
\end{tabular} \\
\begin{tabular}[t]{@{}c@{}}
\includegraphics[width=0.3\textwidth]{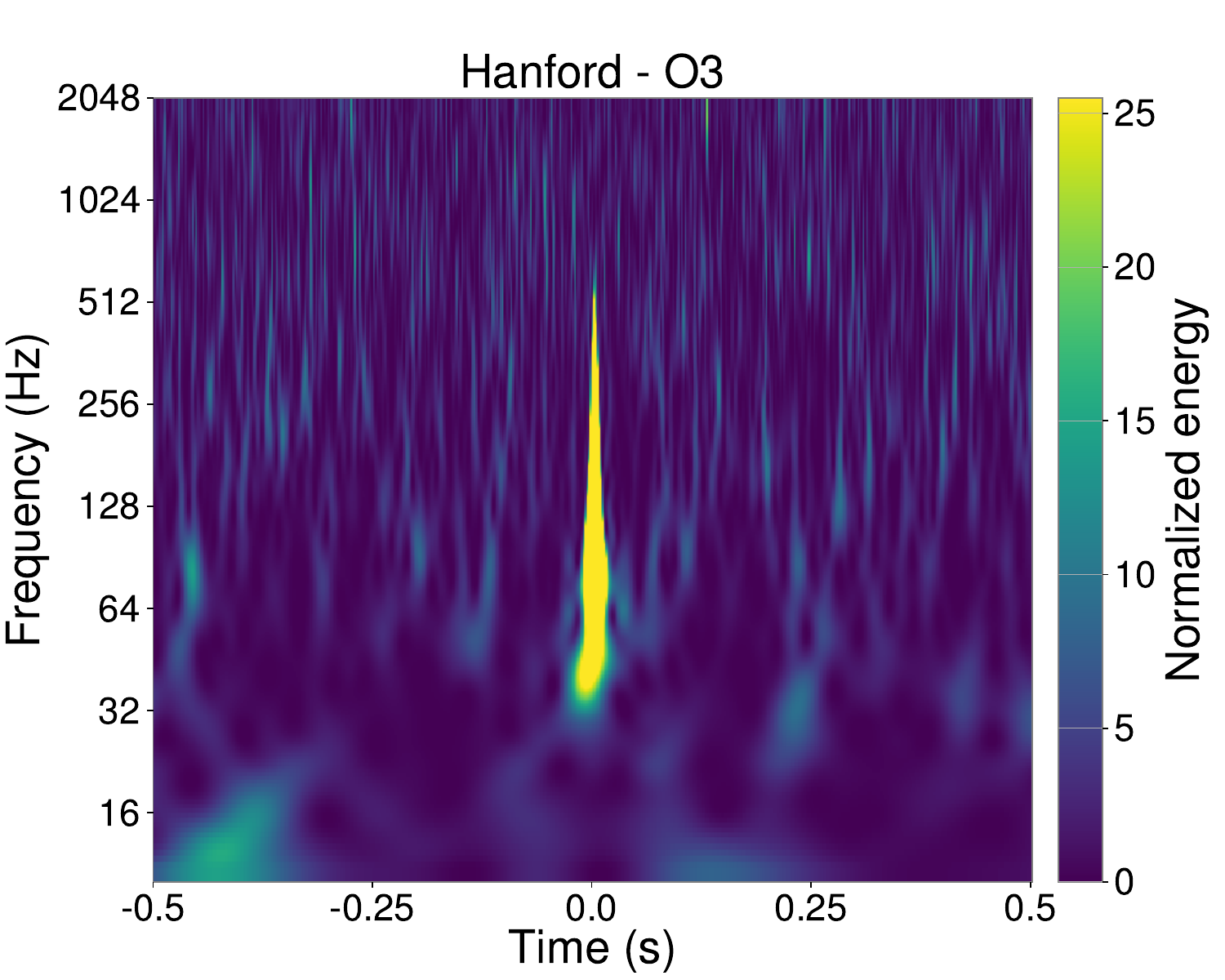} \\
\fontsize{12}{14}\selectfont\textbf{A}
\end{tabular} &
\begin{tabular}[t]{@{}c@{}}
\includegraphics[width=0.3\textwidth]{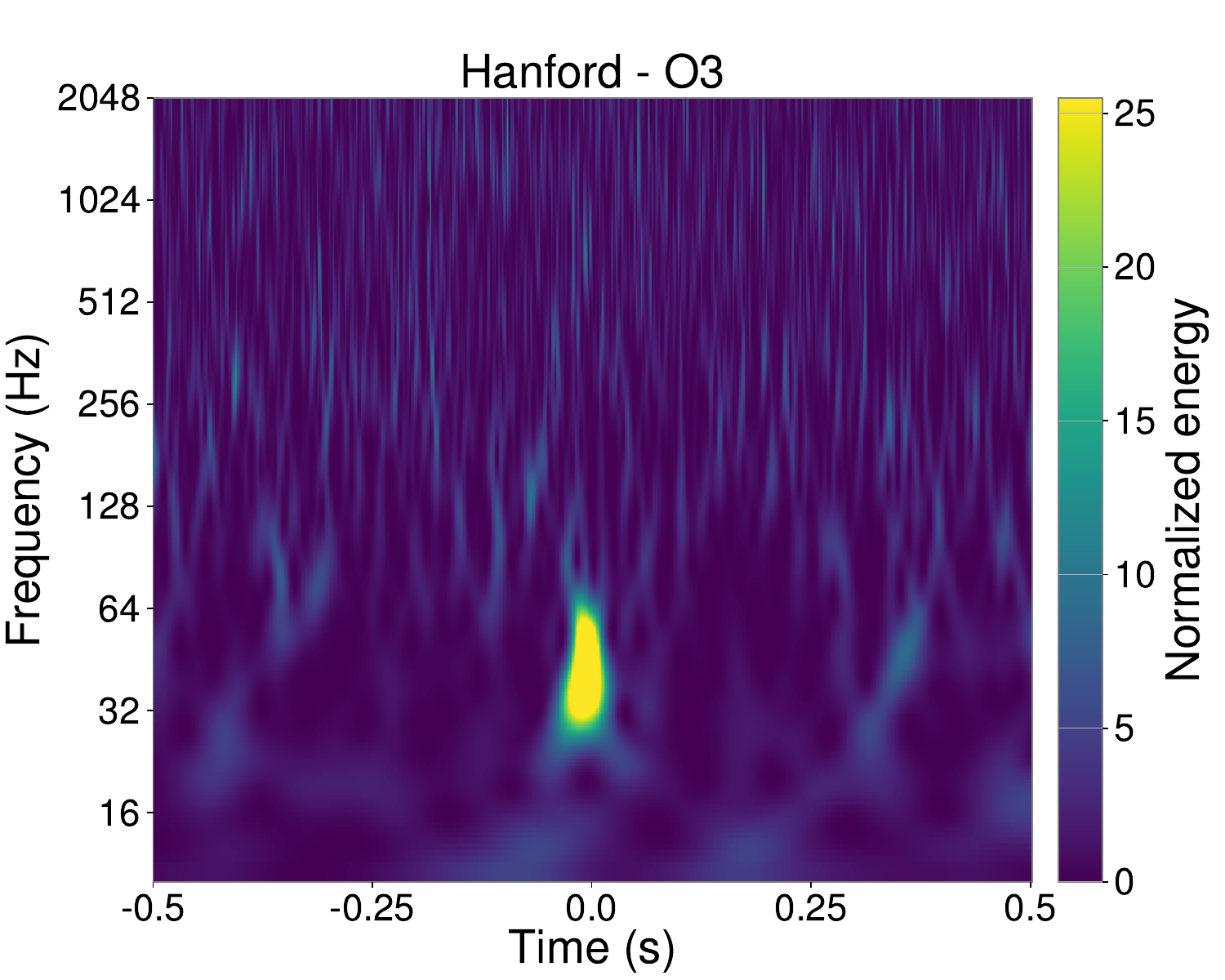} \\
\fontsize{12}{14}\selectfont\textbf{B}
\end{tabular} &
\begin{tabular}[t]{@{}c@{}}
\includegraphics[width=0.3\textwidth]{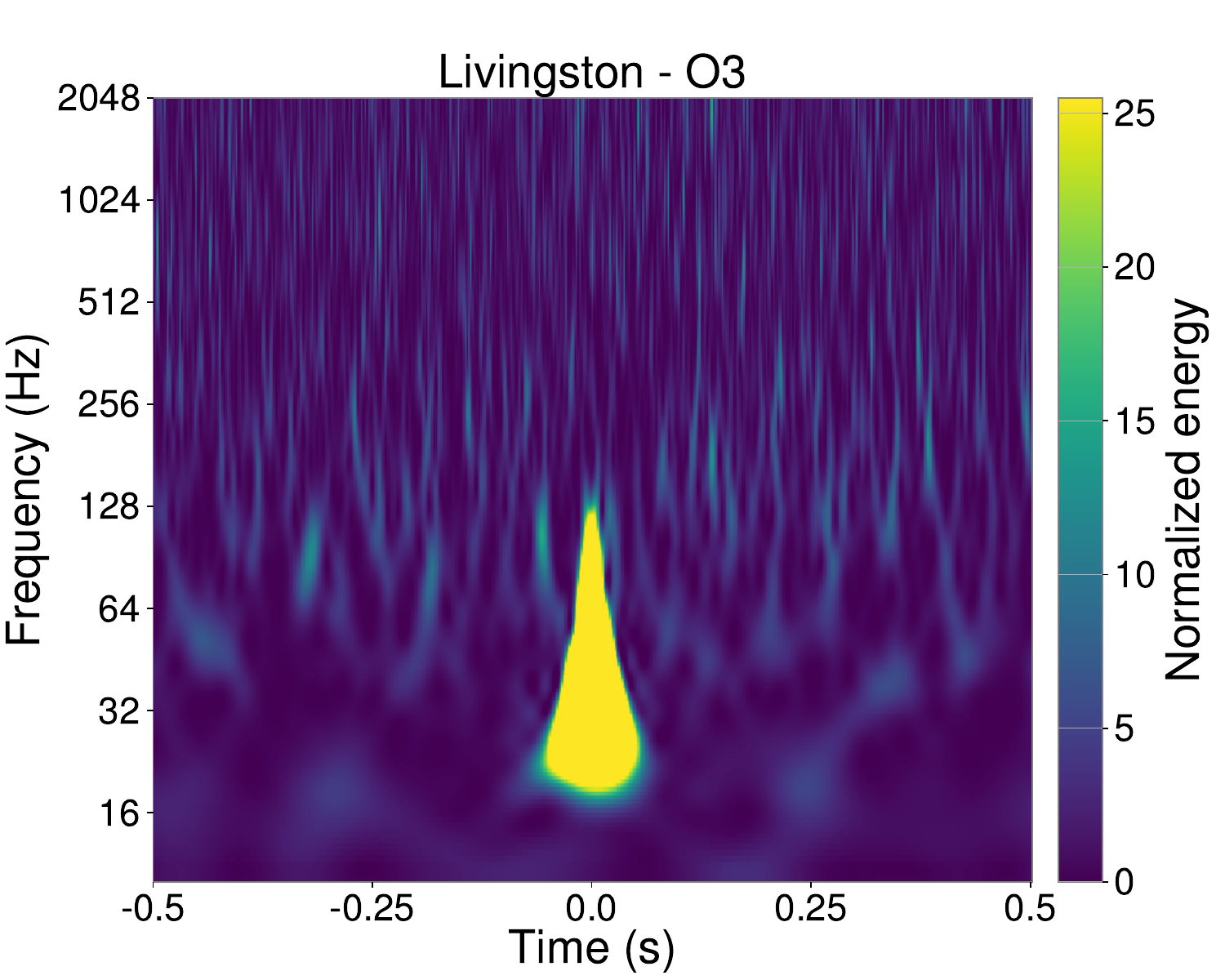} \\
\fontsize{12}{14}\selectfont\textbf{C}
\end{tabular}
\end{tabular}
\caption{Example qscans of high mass CBC GW signals and glitches confused by the original Gravity Spy model. 
A simulated GW source with a total mass of $245~M_{\odot}$ (\textbf{\RNum{1}}) is most commonly mistaken for blips (\textbf{A}) and low frequency blips (\textbf{B}), which share a similar morphology in a qscan. 
In contrast, a simulated signal with $126~M_{\odot}$ total mass and an SNR of 35 (\textbf{\RNum{2}}) is more often confused for a koi fish glitch (\textbf{C}) or tomte (\textbf{D}) glitch. 
While the morphology of these glitches is not as similar, the original Gravity Spy model tended to confuse higher SNR GW for the higher SNR glitch classes. 
}
\label{fig:high_mass_chirp}
\end{figure*}

High mass CBCs are shorter in duration in the sensitive band of the LIGO and Virgo detectors relative to low mass CBCs, as shown in Figure \ref{fig:high_mass_chirp} for a simulated GW sources with total mass $245~M_{\odot}$ (\textbf{\RNum{1}}) and $126~M_{\odot}$ (\textbf{\RNum{2}}). 
These GWs often resemble short duration glitch types shown in Figure \ref{fig:high_mass_chirp}, including blips (\textbf{A}) and low frequency blips (\textbf{B}). 
The similar frequency range and morphology make distinguishing between high mass CBCs and these glitch classes difficult. 
However, high mass CBCs are confused less often for blip glitches compared to low frequency blips, in part due to excess power above $300~\textrm{Hz}$, which high mass GW sources typically do not share (see Figure \ref{fig:high_mass_chirp}).

For the original Gravity Spy model, high mass GW sources were often confused for low frequency blips, and sometimes (in $\sim 3\%$ of cases) blip glitches. This is our main challenge in terms of building a reliable classification system.
In addition, increased GW SNR also led to misclassifications of the higher SNR glitch classes, koi fish (\textbf{C})  and tomtes (\textbf{D}), likely due to the limited representation of louder GW signals in the original Gravity Spy training set. 
Our investigation into louder GW signals found that $\sim50\%$ of misclassifications of high mass CBCs were classified as koi fish or tomte glitches.
As a result, we incorporated higher SNR signals into the training set of our new signal-vs-glitch classifiers. 

\subsection{Extremely high mass CBCs}\label{sec:ex_high_mass}

\begin{figure*}[ht]
\centering
\begin{minipage}[b]{0.45\textwidth}
\centering
\includegraphics[width=\textwidth]{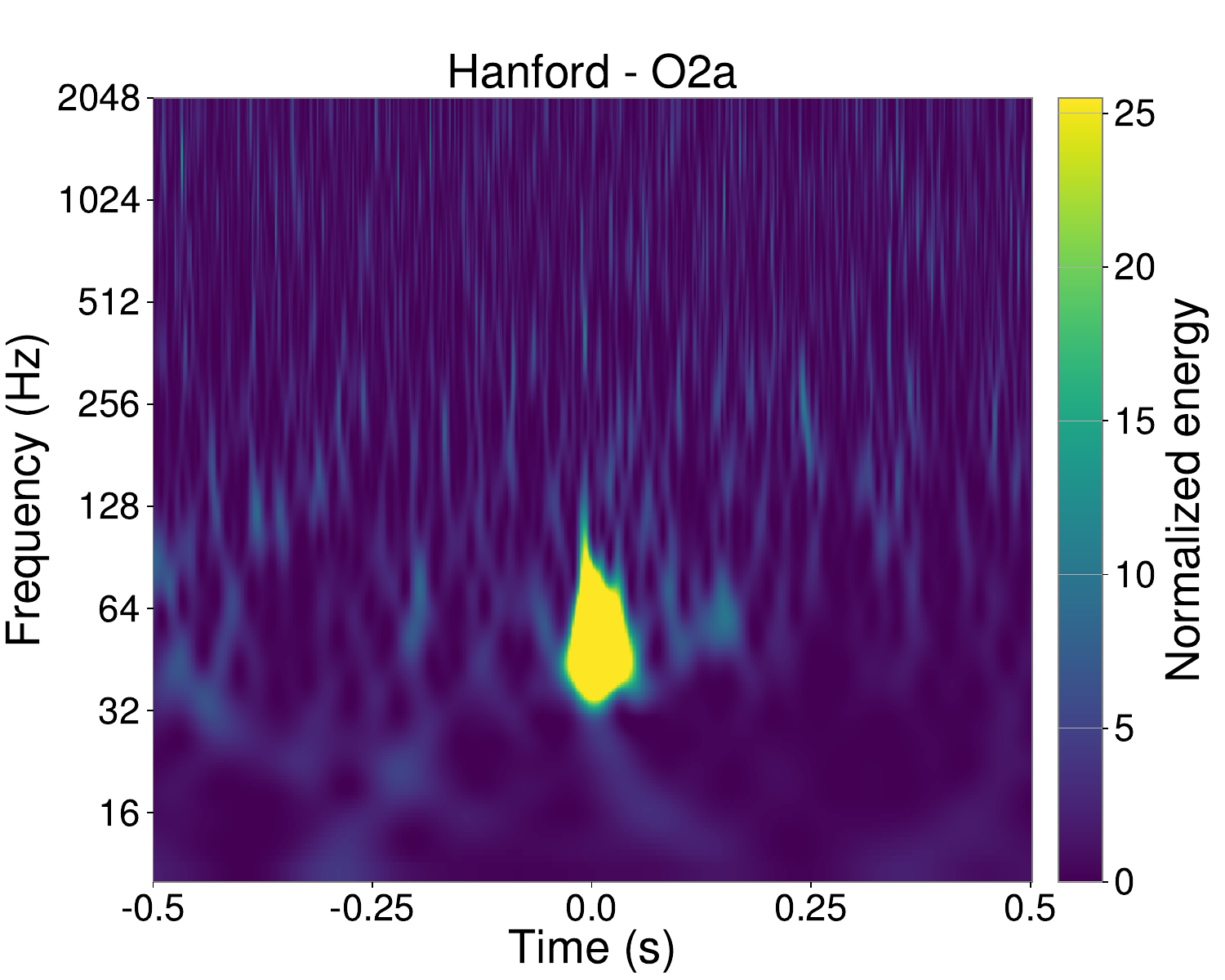}
\end{minipage}
\hspace{0.5cm}
\begin{minipage}[b]{0.45\textwidth}
\centering
\includegraphics[width=\textwidth]{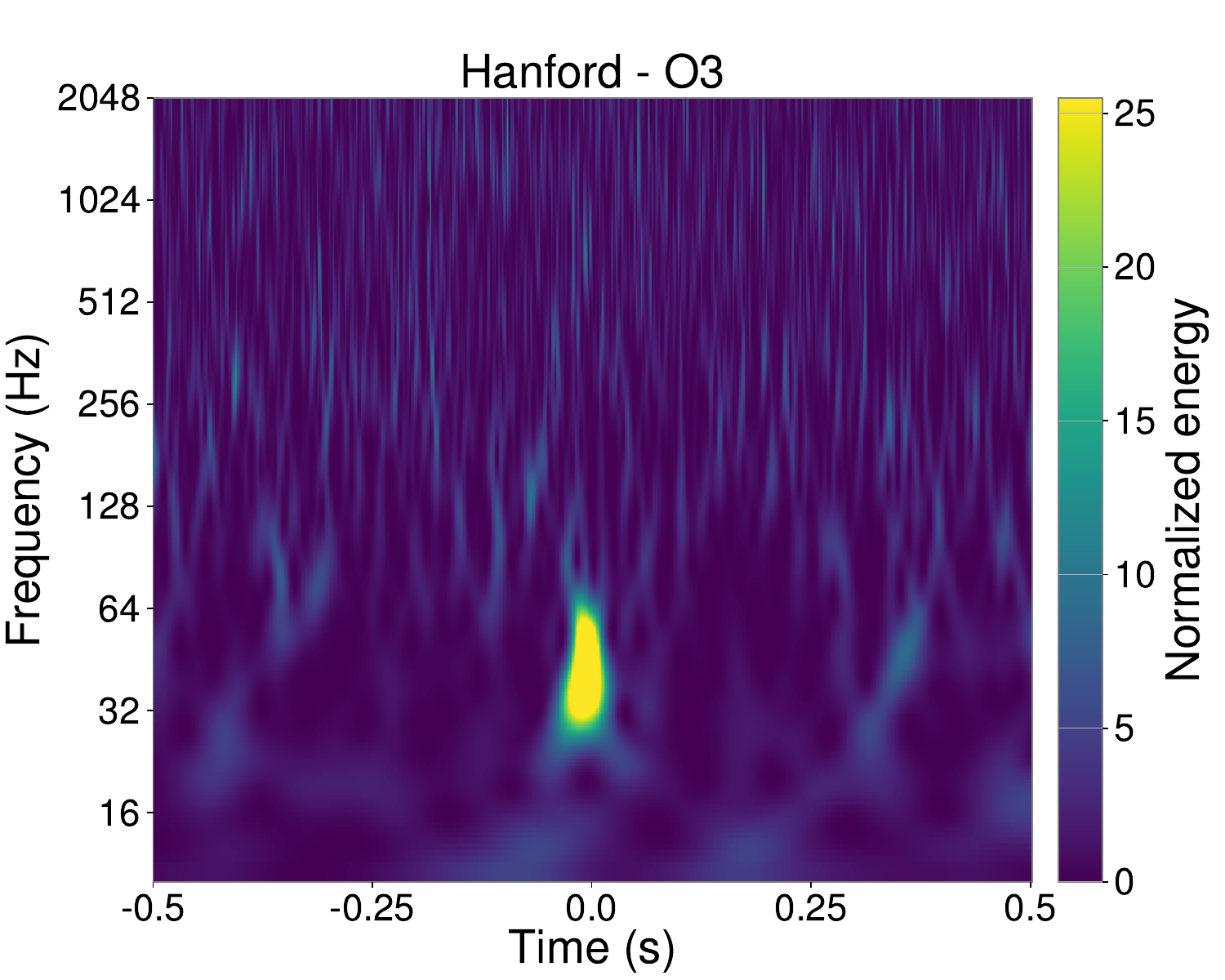}
\end{minipage}
\caption{On the left is a simulated CBC GW signal with a total mass of $281~M_\odot$ and on the right is an example of a low frequency blip glitch. Signals with mass above $250~M_\odot$ overlap very closely with low frequency blip glitches in frequency range and duration, which makes them particularly difficult for image classifiers to differentiate.}
    \label{fig:ex_high_mass_chirp}
\end{figure*}

For mergers beyond $250M_\odot$, the original Gravity Spy model classifies $100\%$ of these signals as low frequency blips with $\sim99\%$ confidence. For example, a $270M_{\odot}$ CBC signal shares strong similarities in duration, frequency range, and morphology with low frequency blips, as shown in Figure \ref{fig:ex_high_mass_chirp}. 
Since the original Gravity Spy model could not classify any CBCs above $250M_\odot$, we investigated these types of signals separately and labeled them as ``extremely high mass" GW sources. 
Further investigation, described in Section \ref{sec:rescale}, suggests that treating this class separately is necessary.

\section{Gravity Spy as a base for a signal-vs-glitch classifier}\label{sec_4}

We first describe our investigations of the original Gravity Spy model's potential to distinguish between GW signals and detector glitches. 
We then discuss our improvements to the original Gravity Spy model for use in this new capacity, targeting the challenges of the classes discussed in the previous section.

\subsection{Testing the original Gravity Spy}\label{Test_GSpy}

To understand the performance of the original Gravity Spy model as a possible signal-vs-glitch classifier, we investigated the model's classification of simulated GWs as a function of the total mass, SNR, and spin of a CBC signal. 
We generated a set of simulated signals and tested whether the model would classify them as chirps; the class label corresponding to the limited number of simulated GWs included in Gravity Spy's original training set. 

The parameters for the simulated GW waveforms used in the test set were drawn uniformly from intervals: $\text{mass}_i\in[1M_\odot,125M_\odot]$, $\text{SNR}\in[3,25]$, and $\text{spin}_i\in[-0.95,0.95]$, where $i\in\{1,2\}$ corresponds to each compact object in the simulated merger. We split the simulated signals into the low mass ($3-50~M_\odot$) and high mass ($50-250~M_\odot$) categories such that there were 500 examples in each mass range. 

\begin{figure*}[ht]
\centering
\begin{tabular}{cc}
\begin{tabular}{@{}c@{}}
  \includegraphics[height=0.25\textheight, width=0.5\textwidth, keepaspectratio]{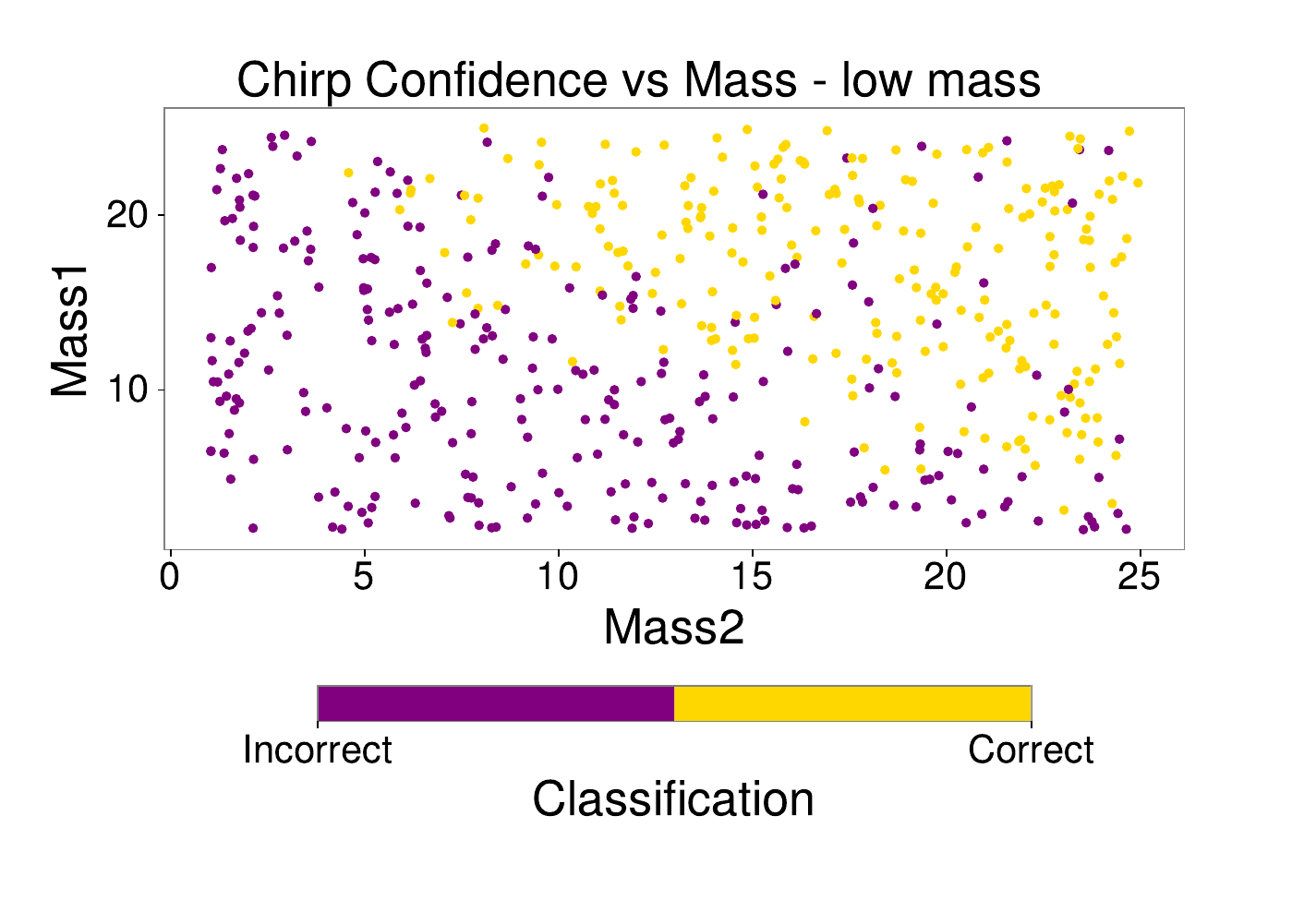} \\
  \textbf{A}
\end{tabular} &
\begin{tabular}{@{}c@{}}
  \includegraphics[height=0.25\textheight, width=0.5\textwidth, keepaspectratio]{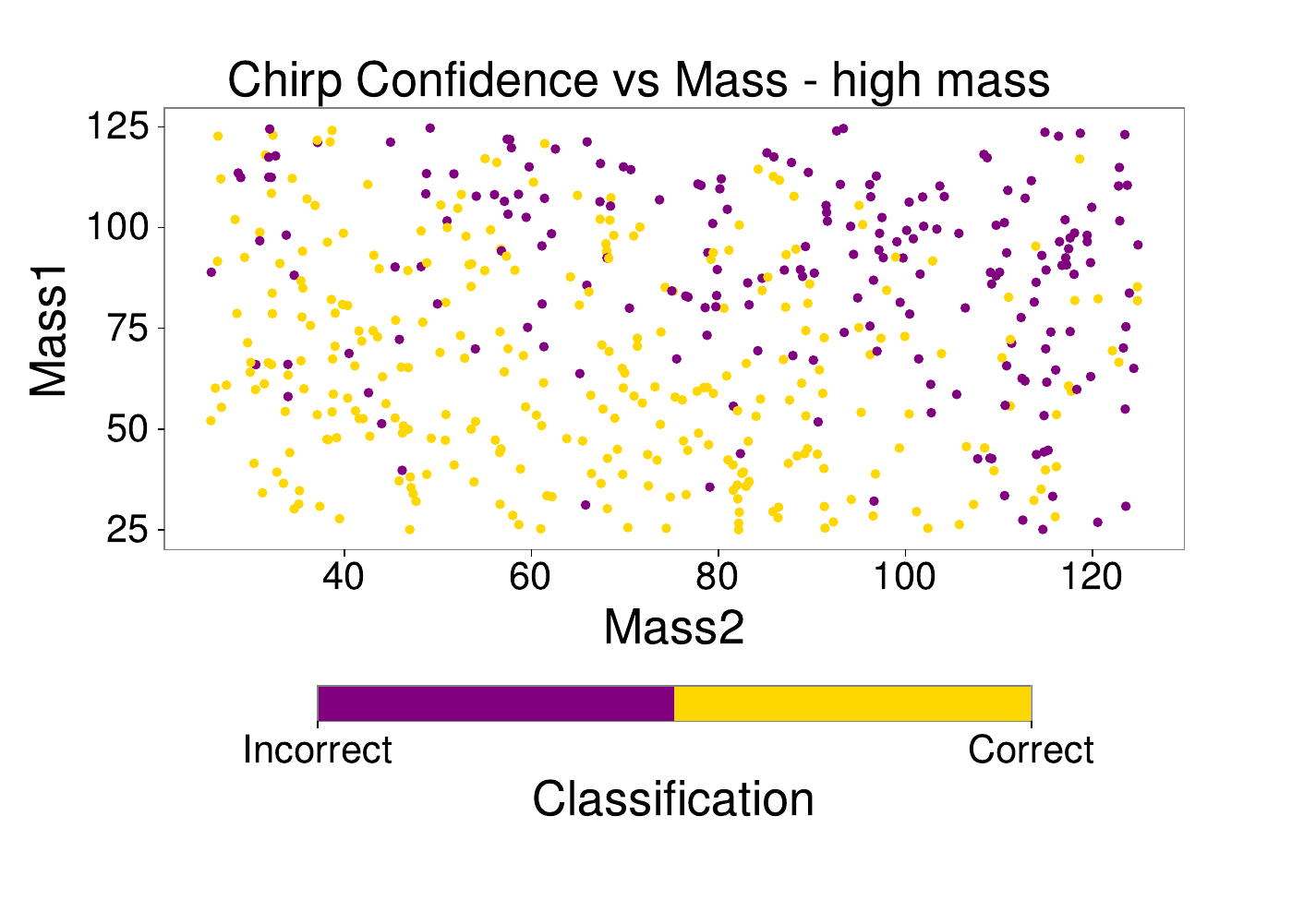} \\
  \textbf{B}
\end{tabular}
\end{tabular}
\caption{The original Gravity Spy model's classification of $500$ simulated signals. The left plot (\textbf{A}) shows low mass sources with parameters $\text{mass}_i\in[1M_\odot,25M_\odot]$ where $i\in\{1,2\}$ corresponds to each binary component in the simulated merger, $\text{SNR}\in[3,25]$, and $\text{spin}_i\in[-0.95,0.95]$. The right plot (\textbf{B}) shows high mass sources with parameters $\text{mass}_i\in[25M_\odot,125M_\odot]$ with the same spin and SNR ranges as in (\textbf{A}). 
"Chirps" that have been correctly classified by the original Gravity Spy model are colored yellow, and incorrect classifications are purple. 
While the simulated GWs have a range of SNR and spin values, the regions where the original Gravity Spy model is able to perform best are clearly dominated by the source's total mass, motivating a method that targets the corresponding morphological differences.}
\label{fig:mass_range_plots}
\end{figure*}

Figure \ref{fig:mass_range_plots} shows the original Gravity Spy's classification\footnote{Note that while Gravity Spy produces confidence ratings for each class a GW signal or glitch could be classified as, our investigation was only interested in Gravity Spy's ability to differentiate between GWs and glitches. So, in our analysis of Gravity Spy classifications, only the classification with the highest confidence was considered.} of these simulated signals, plotted by the component masses $\text{mass}_i$, with classification accuracy shown as color. 
Of the $273$ false negatives in the low mass range (Figure \ref{fig:mass_range_plots} \textbf{A}), $81\%$ were classified as a scratchy glitch while $17\%$ were classified as Gaussian noise (`no glitch'). 
We note again that for the purposes of this study, a no glitch classification is equivalent to a GW classification, as both classes are not actionable for further data quality studies in GW event candidate validation. 
The misclassifications of higher mass simulated GW signals (Figure \ref{fig:mass_range_plots} \textbf{B}) were slightly more diverse with the majority of the $204$ false negatives being classified as either a blip or low frequency blip.

These results are evidence that using the original Gravity Spy's CNN outside of its original intended purpose is not sufficient for an accurate signal-vs-glitch classifier without further modification, as we require high accuracy for automation for GW candidate event validation. 
In particular, GWs require more representation in our training set relative to the original Gravity Spy training set. 
Our results are consistent with the findings reported in Bahaadini et al.~\cite{BAHAADINI2018172}, which identified a higher rate of inaccuracies associated with signals that had a poorer representation in the original Gravity Spy training set.  

Only 60 out of the 9631 original Gravity Spy training set examples were simulated GWs, as shown in Table \ref{tab:trn_set}.
These simulated GWs were labeled as ``chirps" to also capture potential transient noise increasing in frequency over time with a similar morphology~\cite{GSpy}. 
Given the lack of GW examples, we might expect few correct classifications of a broad range of simulated GWs as ``chirps". 
However $79\%$ of GW sources with a total mass within $25M_{\odot}$ and $150M_{\odot}$ were correctly classified by the original Gravity Spy model. 

\begin{table*}[!htb]
\begin{tabular}{| p{3.45cm} | p{3.45cm} | p{3.45cm} | p{3.45cm} |}
    \hline
    Glitch & Training Examples & Glitch & Training Examples\\ \hline
    Blip & 1821 & Koi Fish & 706 \\ \hline
    Tomte & 703 & LF Blip & 630 \\ \hline
    LF Burst & 621 & Scattered Light & 593\\ \hline
    Light Modulation & 512 & Power Line & 449\\ \hline
    LF Lines & 447 & Extremely Loud & 447\\ \hline
    Violin Mode & 412 & Fast Scattering & 400\\ \hline
    Scratchy & 337 & 1080 Lines & 327\\ \hline
    Whistle & 299 & Helix & 279\\ \hline
    Repeating Blips & 263 & No Glitch & 117\\ \hline
    1400 Ripples & 81 & Chirp & 60\\ \hline
    Air Compressor & 58 & Wandering line & 42\\ \hline
    Paired Doves & 279 & & \\ \hline
    \end{tabular}\par
 \caption{Gravity Spy's original training set~\cite{Glanzer:2022avx}. Of the 9631 training examples in the original training set, the number of simulated GW (``chirp") examples was particularly small, as GW classification was outside of Gravity Spy's original intended use.}
\label{tab:trn_set}
\end{table*}

In summary, we identified a clear relation between total mass and the original Gravity Spy model's classification accuracy of GWs, motivating splitting up the classifier to target different mass ranges.  
Additionally, we showed that only a small subset of glitch classes cause GW misclassifications in the original Gravity Spy model, which is evidence that 20 glitch classes are unnecessary for reliable signal versus glitch classification and may be introducing confusion for this targeted application. 
These results suggest that better results for signal-vs-glitch classification are attainable with an augmented training set and a restructuring of the Gravity Spy classification model. 

\subsection{Rebuilding Gravity Spy}

We propose a restructuring of the original Gravity Spy classification framework for use in signal-vs-glitch classification. 
Our proposed model consists of two classifiers focused on signals from either end of the mass spectrum and the corresponding glitch classes that we have found the original Gravity Spy model confuses for signals. 

We will call the signal versus glitch model that focuses on low mass CBC sources and similar glitches GSpySVG\_LM and the one that focuses on high mass CBCs and similar glitches GSpySVG\_HM.
In our prototype of this method we supply a balanced training set for each model, reduce the glitch classes to only those that the original Gravity Spy model confused with signals in the corresponding mass range, and increased the diversity of GW morphology included in the GW class to better represent expected GW signals. 

\begin{table*}[!htb]

    \begin{minipage}{.5\linewidth}
      \centering
      \begin{center}
          \hspace{-45px} GSpySVG\_LM
      \end{center}

      \centering

      \hspace{-150px}
        \begin{tabular}{@{} rrc|ccccccc }
\rowcolor{white!30} \cellcolor{white}
   & & \multicolumn{3}{c}{\# of Examples}
  \\
        \cmidrule{2-5}
\rowcolor{black!15} \cellcolor{white}
   & Blip   &&& 150 \\
   & LF Blip     &&& 150   \\
\rowcolor{black!15} \cellcolor{white}
   & No Glitch    &&& 115    \\
   & Scratchy 
                  &&& 150    \\
\rowcolor{black!15} \cellcolor{white}
 \rotnine{\rlap{~True Class}}
   & LM GW     &&& 150   \\
        \cmidrule[1pt]{2-7}
    \end{tabular}
    \end{minipage}%
        \begin{minipage}{.5\linewidth}
      \centering
      \begin{center}
          \hspace{-45px} GSpySVG\_HM
      \end{center}

      \centering

      \hspace{-150px}
        \begin{tabular}{@{} rrc|ccccccc }
   % & & \multicolumn{3}{c}{GSpySVG\_LM} \\[2ex]
\rowcolor{white!30} \cellcolor{white}
   & & \multicolumn{3}{c}{\# of Examples}
  \\
        \cmidrule{2-5}
\rowcolor{black!15} \cellcolor{white}
   & Blip   &&& 150 \\
   & LF Blip     &&& 150   \\
\rowcolor{black!15} \cellcolor{white}
   & Koi Fish    &&& 150    \\
   & Tomte 
                  &&& 150    \\
\rowcolor{black!15} \cellcolor{white}
 \rotnine{\rlap{~True Class}}
   & HM GW     &&& 150   \\
        \cmidrule[1pt]{2-7}
    \end{tabular}
    \end{minipage}%
    \caption{Our new method consists of two classifiers based on the Gravity Spy model. GSpySVG\_LM is trained to distinguish between low mass signals and similar glitches and GSpySVG\_HM is trained to handle high mass signals and similar glitches. 
    We list the glitch classes used in each classifier as well as the number of examples used in the corresponding training set. 
    Although this prototype method uses fewer training examples, it has a more equal representation of GWs relative to glitch classes compared to the original Gravity Spy training set. 
   We only included glitch classes that Gravity Spy's original model confused with either low mass or high mass GW signals.}
\end{table*}

\begin{figure*}[ht]
\centering
\begin{tabular}{cc}
\begin{tabular}{@{}c@{}}
  \includegraphics[height=0.25\textheight, width=0.5\textwidth, keepaspectratio]{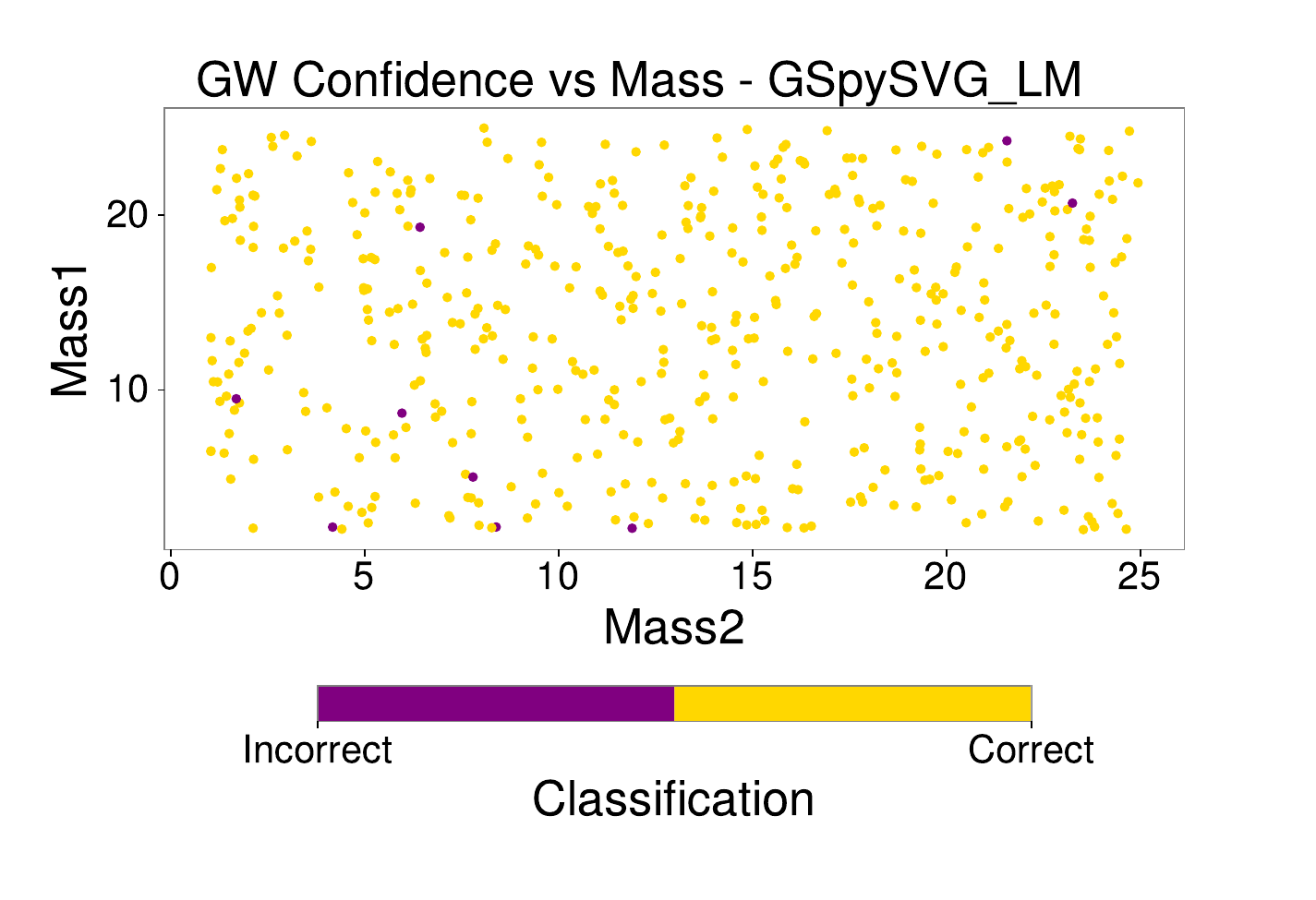} \\
  \textbf{A}
\end{tabular} &
\begin{tabular}{@{}c@{}}
  \includegraphics[height=0.25\textheight, width=0.5\textwidth, keepaspectratio]{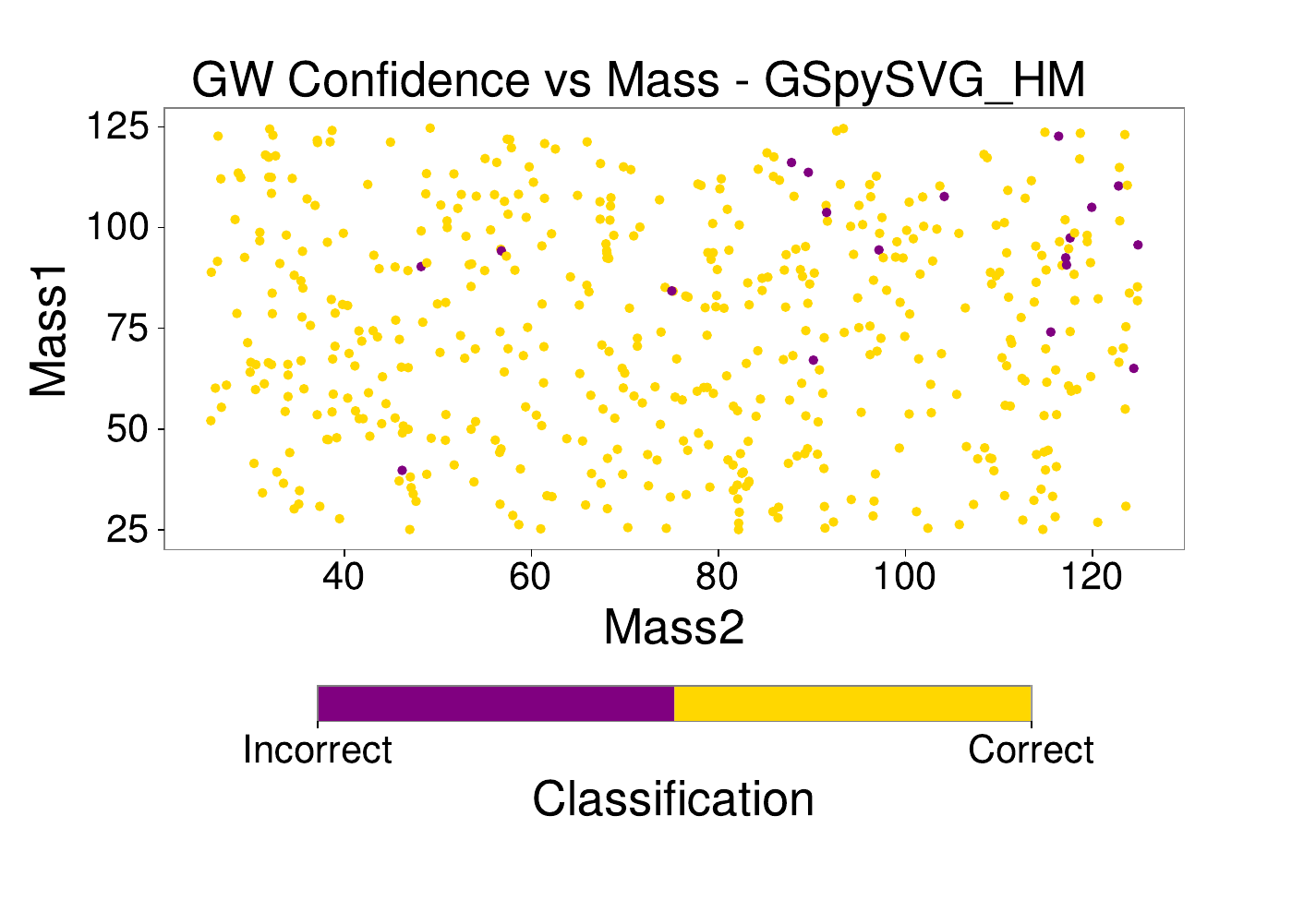} \\
  \textbf{B}
\end{tabular}
\end{tabular}
\caption{GSpySVG\_LM (\textbf{A}) and GSpySVG\_HM (\textbf{B}) tested on the same test set and mass range used for the original Gravity Spy model, as reported in Figure \ref{fig:mass_range_plots}.
As in Figure \ref{fig:mass_range_plots}, the data is presented with the mass values on the $x$ and $y$ axes, while signals that have been correctly identified are colored yellow and incorrect classifications are purple. 
Compared with the original Gravity Spy model, our prototype produces a substantial decrease in GW misclassifications: 273 to 9 for simulated signals with total mass in the range $3-50 M_\odot$ and 204 to 19 for simulated signals with total mass in the range $50-300 M_\odot$.}
\label{fig:results}
\end{figure*}

After we implemented these changes, we found results significantly improved compared with the original Gravity Spy model. 
Using the same tests sets and mass ranges as reported for the original Gravity Spy model in section \ref{Test_GSpy} Figure~\ref{fig:mass_range_plots}, we evaluated our new method's ability to correctly classify simulated GW signals, as shown in Figure~\ref{fig:results}. 
Most notably, the number of simulated signals in the $3-250M_\odot$ range classified correctly is much higher for GSpySVG\_HM and GSpySVG\_LM compared with the performance of original Gravity Spy, as shown in Figures \ref{fig:mass_range_plots} and \ref{fig:results} respectively. 
While the original Gravity Spy model was only able to classify $52\%$ of simulated signals in the same mass range correctly, GSpySVG\_HM and GSpySVG\_LM were able to classify $97\%$ of them correctly.

We also observed similarly accurate results for glitch classifications by GSpySVG\_HM and GSpySVG\_LM. We found that both GSpySVG\_HM and GSpySVG\_LM were able to correctly classify $99\%$ of the $200$ test glitches as belonging to the correct glitch class. 
This shows our method is able to increase signal classification accuracy while maintaining strong performance in glitch classification. We note, however, that novel glitch classes that could appear in future observing runs were not considered as the focus of this proof-of-principle study was to demonstrate feasibility. For a study on the robustness of a wider range of glitch classes we refer the reader to the follow-up paper by Alvarez et al \cite{alvarezlopez2023gspynettree}.

\begin{table*}[!htb]
    \begin{minipage}{.5\linewidth}
      \centering

    GSpySVG\_LM Classification

        \begin{tabular}{@{} cr*{10}c }
   % & & \multicolumn{10}{c}{GSpySVG\_LM Classification} \\[2ex]
\rowcolor{white!30} \cellcolor{white}
   & & \rot{Blip} & \rot{LF Blip} & \rot{No Glitch} & \rot{Scratchy} 
   & \rot{GW} \\
        \cmidrule{2-7}
\rowcolor{black!15} \cellcolor{white}
   & Blip   &\hl{50} & 0   &  0  &  0  & 0     \\
   & LF Blip     & 0 & \hl{50}   &  0  &  0  & 0     \\
\rowcolor{black!15} \cellcolor{white}
   & No Glitch    & 0 & 0   &  \hl{49}  &  1  & 0      \\
   & Scratchy 
                  & 0 &  0  &  0  &  \hl{48}  & 2    \\
\rowcolor{black!15} \cellcolor{white}
 \rotnine{\rlap{~True Class}}
   & GW     & 0 &  0  &  8  &  1  & \hl{491}     \\
        \cmidrule[1pt]{2-7}
    \end{tabular}
    \end{minipage}%
    \begin{minipage}{.5\linewidth}
      \centering

GSpySVG\_HM Classification
      
        % \caption{}
    \begin{tabular}{@{} cr*{10}c }
   % & & \multicolumn{10}{c}{GSpySVG\_HM Classification} \\[2ex]
\rowcolor{white!30} \cellcolor{white}
   & & \rot{Blip} & \rot{LF Blip} & \rot{Koi Fish\hspace{.2cm}} & \rot{Tomte} 
   & \rot{GW} \\
        \cmidrule{2-7}
\rowcolor{black!15} \cellcolor{white}
   & Blip   &\hl{50} & 0   &  0  &  0  & 0     \\
   & LF Blip     & 0 & \hl{50}   &  0  &  0  & 0     \\
\rowcolor{black!15} \cellcolor{white}
   & Koi Fish    & 0 & 0   &  \hl{47}  &  3  & 0      \\
   & Tomte 
                  & 0 &  0  &  0  &  \hl{50}  & 0    \\
\rowcolor{black!15} \cellcolor{white}
 \rotnine{\rlap{~True Class}}
   & GW     & 4 &  15  &  0  &  0  & \hl{481}     \\
        \cmidrule[1pt]{2-7}
    \end{tabular}
    \end{minipage} 
    \caption{A confusion matrix for GSpySVG\_LM (left) and GSpySVG\_HM (right) performance, as shown in Figure~\ref{fig:results}. 
    Both GSpySVG\_LM and GSpySVG\_HM correctly classified $99\%$ of the test glitches while GW classification accuracy was 98$\%$ and 96$\%$ respectively. 
    Moreover, below a total mass of $200M_{\odot}$, GSpySVG\_HM was able to classify $98\%$ of simulated signals correctly.}
\end{table*}

Further testing showed GSpySVG\_LM and GSpySVG\_HM correctly classified $75\%$ of the retracted O3b GW event candidates~\cite{gwtc3,gwtc21} as non-astrophysical. 
Additionally, we tested the models on all confirmed O3b candidate events~\cite{gwtc3}. GSpySVG\_HM was able to correctly classify all 10 signals that fell under the high mass GW class and GSpySVG\_LM correctly classified 9/11. 
The two O3b confirmed GW candidates misclassified by GSpySVG\_LM were both classified as no glitch. So, of all 21 signals in O3b, none would have been flagged as a glitch by these new models. 

\subsection{Offsets}

We also considered glitches and GW signals with a slight time offset relative to the reported merger time of the candidate. 
We confirmed that, as previously reported, shifting signals with respect to the center of the qscan results in a significant decrease in Gravity Spy classification performance. 
In particular, we saw a $16\%$ decrease in signal classification accuracy when we introduced random offsets in the range of $[-0.1,0.1]$ seconds to account for possible offset candidate event times reported in the search pipelines. 

\begin{figure*}[ht]
\centering
\includegraphics[width=.3\textwidth]{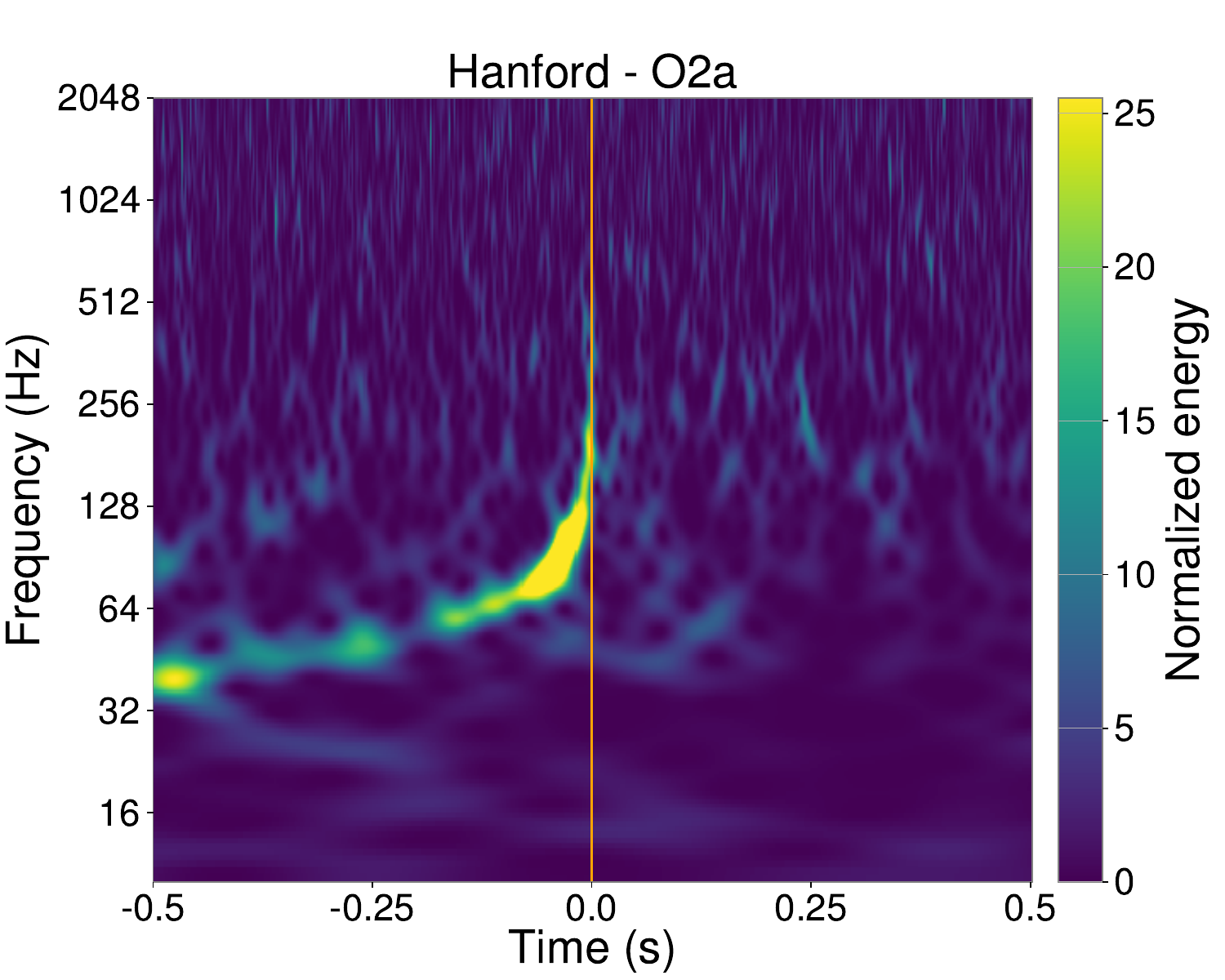}\quad
\includegraphics[width=.3\textwidth]{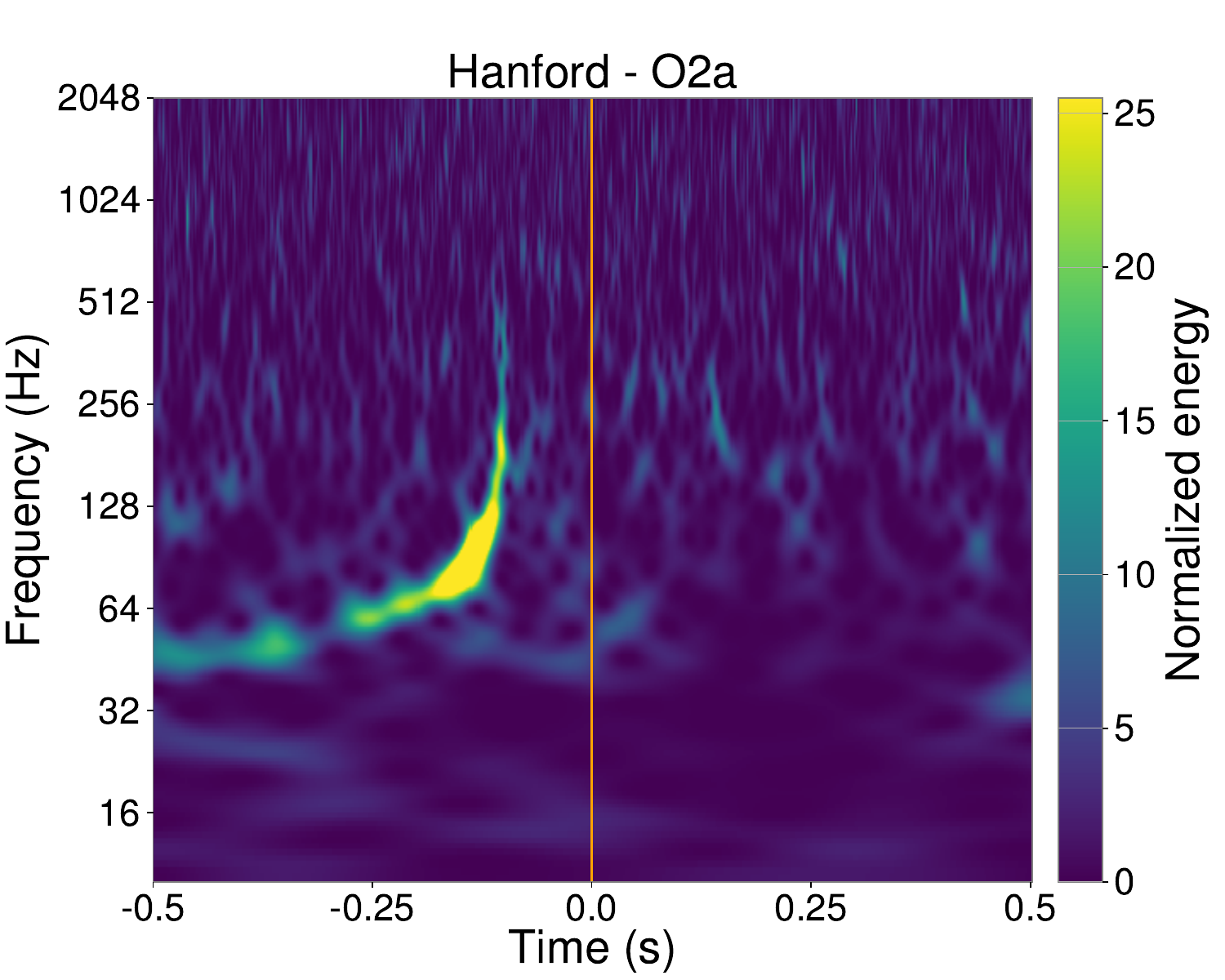}\quad
\includegraphics[width=.3\textwidth]{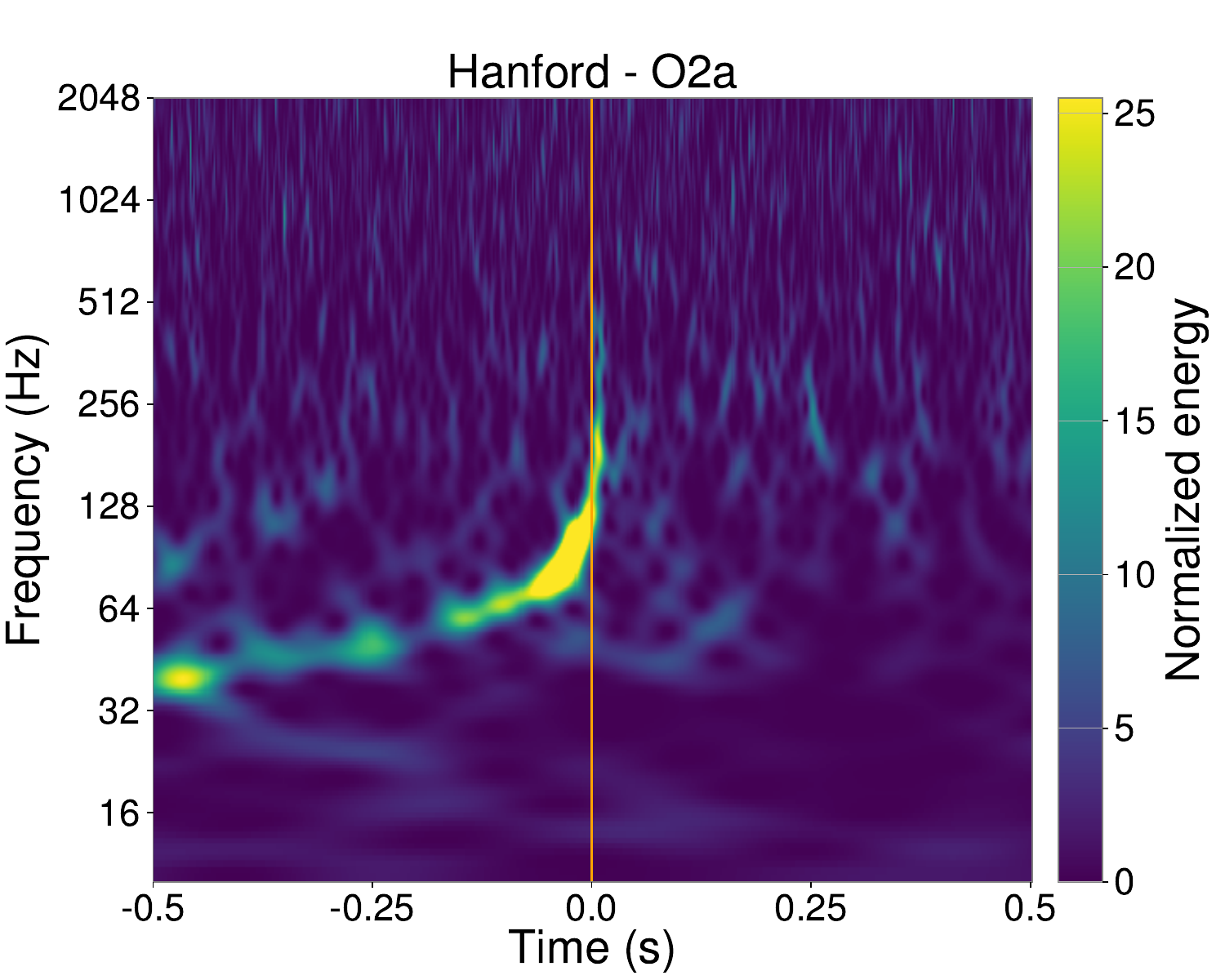}

\medskip

\includegraphics[width=.3\textwidth]{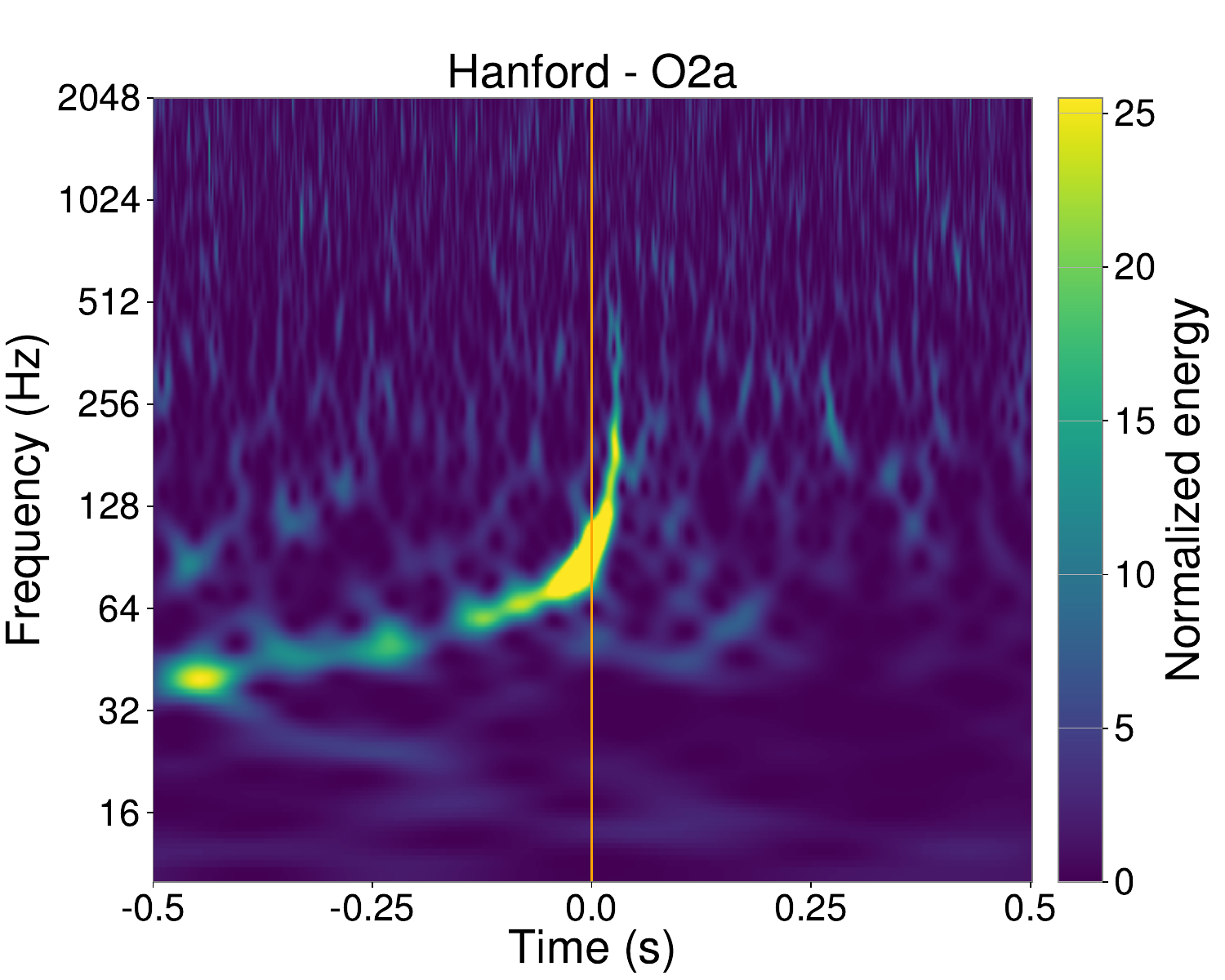}\quad
\includegraphics[width=.3\textwidth]{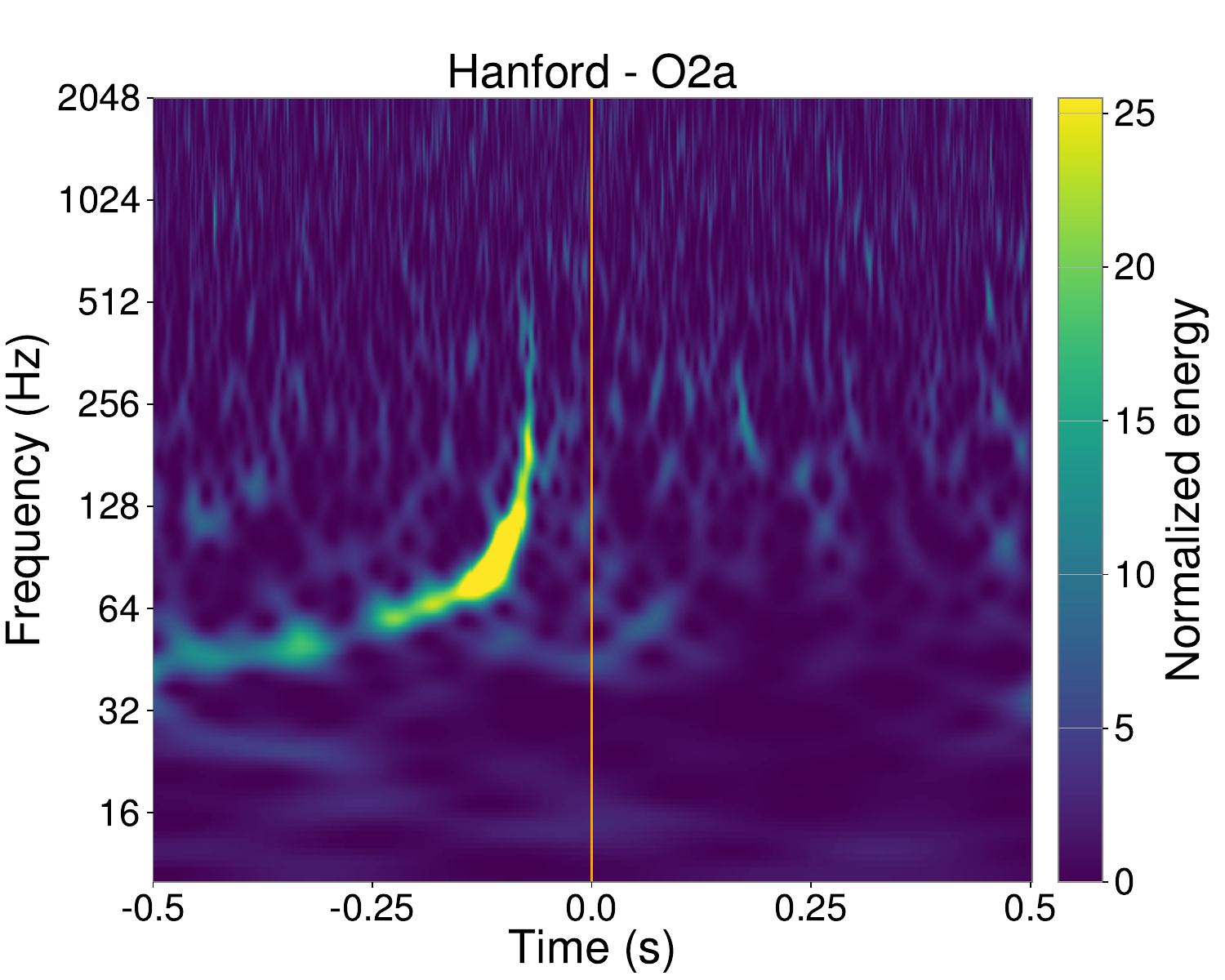}

\caption{An example of a simulated signal with random time offsets. A simulated GW with a merger time at $t=0$ is shown on the top left. The other images are $4$ copies, each containing a random time offset in the range of $[-0.1,0.1]$ seconds. 
We added red center lines at $t=0$ to the images included in this figure as a visual aid.}
    \label{fig:offset}
\end{figure*}

To improve robustness against time offsets for GW event candidates, we supplemented our new training set for the GSpySVG\_LM and GSpySVG\_HM models with time-translated examples. 
We made $4$ copies of each example in our GSpySVG\_LM and GSpySVG\_HM training sets and added a random offset in the range of $[-0.1,0.1]$ seconds to each of the copies. 
This supplemented training set increased classification accuracy from $83\%$ to $99\%$ on the same test set of time-translated images. 
We saw a negligible difference in performance on the un-translated test set. Figure \ref{fig:offset} gives an example of a simulated GW signal in our original training set and $4$ copies with random time offsets, as used in our supplemented training set.

\subsection{Re-scaled qscans for CBC sources with total mass above $250M_\odot$}\label{sec:rescale}

In section \ref{sec:ex_high_mass}, we noted that signals with a total mass above $250M_\odot$ were extremely challenging for the original Gravity Spy CNN to classify. 
When testing the original Gravity Spy model on a test set of 50 simulated signals with total masses between $250M_\odot$ and $350M_\odot$, none of them were classified correctly. Even after retraining for signals in the $50$ to $250M_\odot$ range, GSpySVG\_HM was only able to correctly classify $\sim8\%$ of these signals. 
Both the original Gravity Spy model and GSpySVG\_HM misclassified these simulated signals as low frequency blip glitches.

In this mass range, signals and low frequency blip glitches are nearly indistinguishable by eye and with a CNN using qscans as a feature set, as shown in Figure \ref{fig:ex_high_mass_chirp}. 
To overcome this challenge, we introduce a new feature set using qscan processing techniques that give our method more distinguishing power.

\begin{figure*}[ht]
\centering
\begin{tabular}{cc}
\begin{tabular}{@{}c@{}}
  \includegraphics[height=0.5\textheight, width=0.5\textwidth, keepaspectratio]{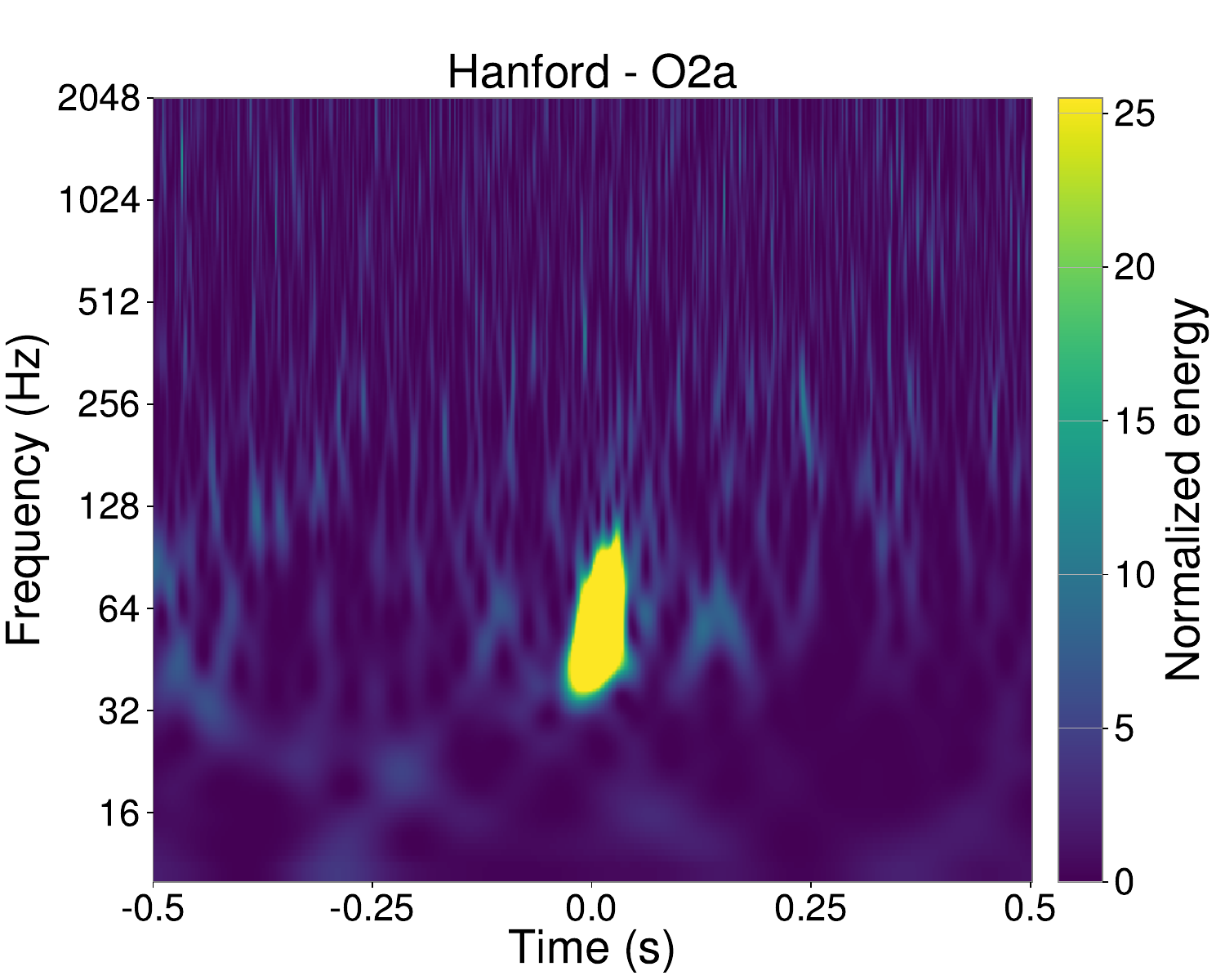} \\
  \textbf{A}
\end{tabular} &
\begin{tabular}{@{}c@{}}
  \includegraphics[height=0.5\textheight, width=0.5\textwidth, keepaspectratio]{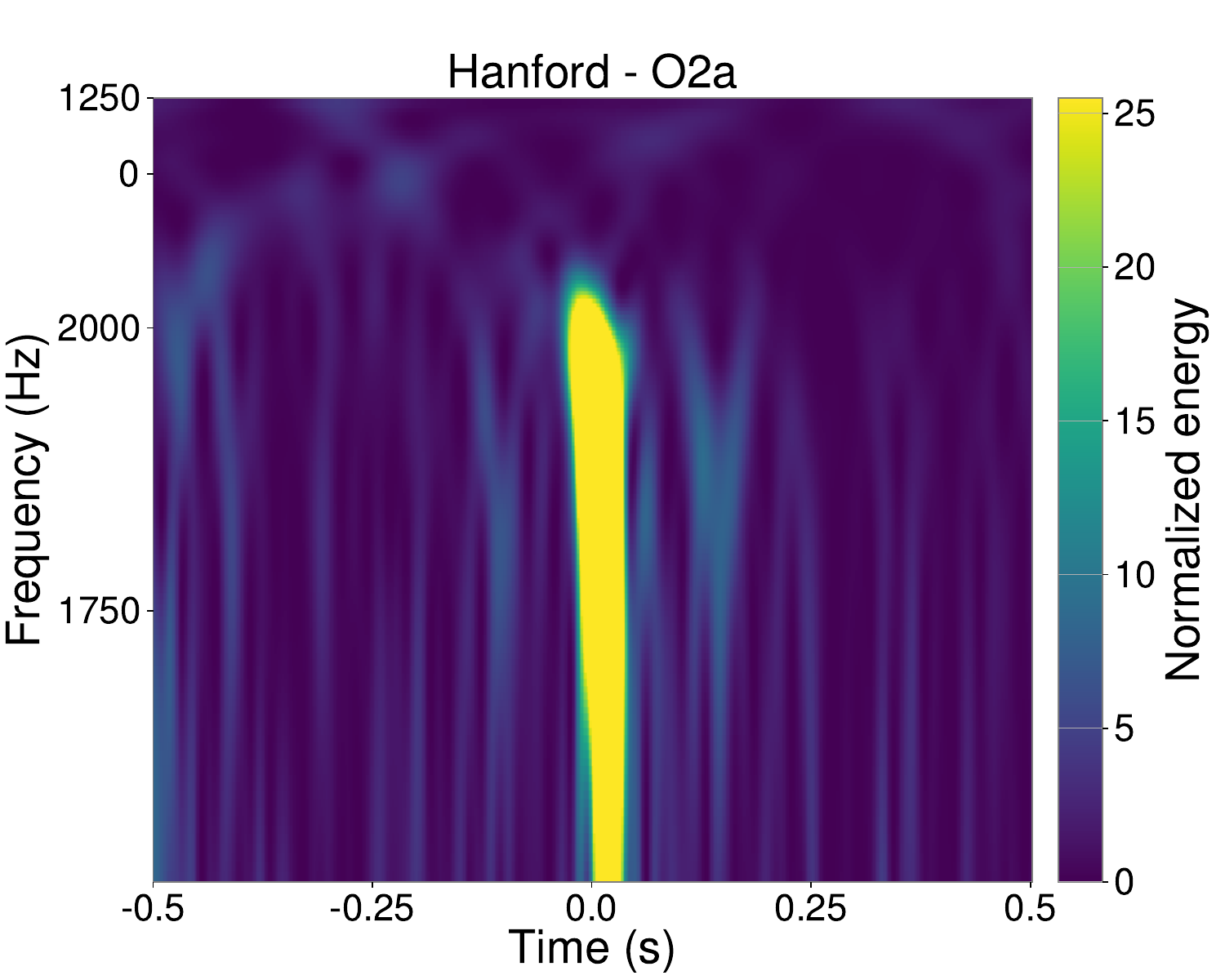} \\
  \textbf{B}
\end{tabular} \\
\begin{tabular}{@{}c@{}}
  \includegraphics[height=0.5\textheight, width=0.5\textwidth, keepaspectratio]{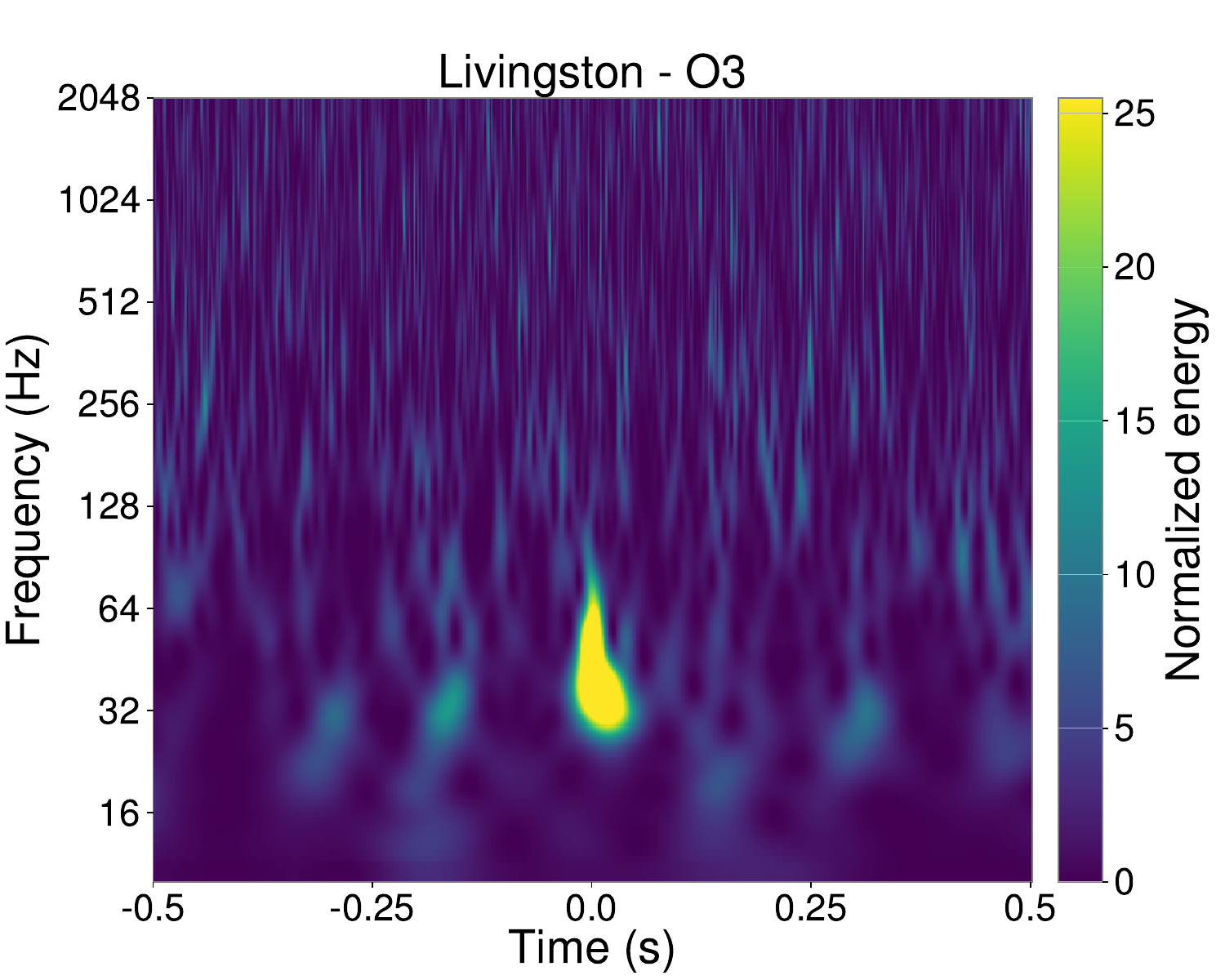} \\
  \textbf{C}
\end{tabular} &
\begin{tabular}{@{}c@{}}
  \includegraphics[height=0.5\textheight, width=0.5\textwidth, keepaspectratio]{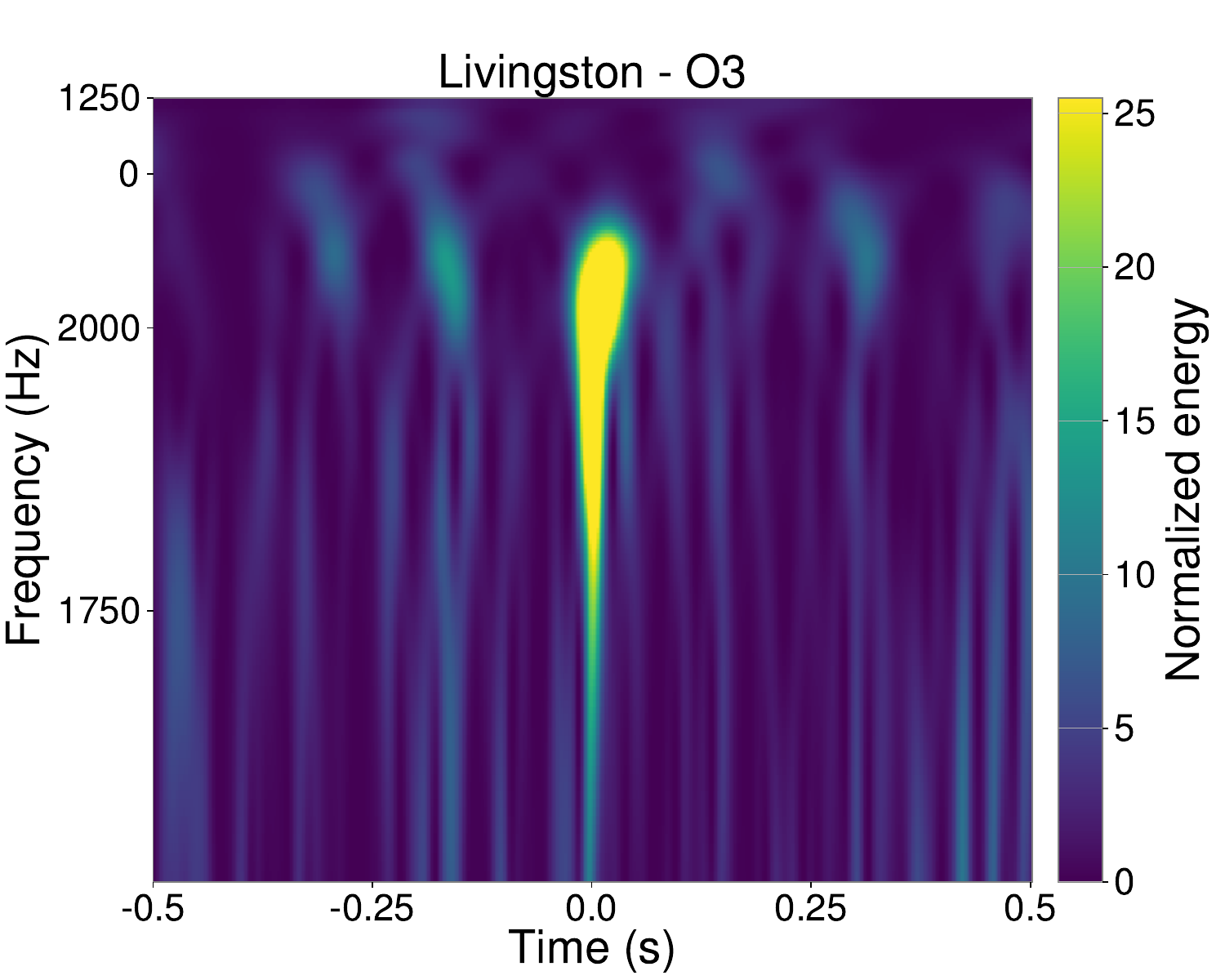} \\
  \textbf{D}
\end{tabular}
\end{tabular}
\caption{An example of a simulated $270M_\odot$ merger (\textbf{A}) and the same signal after applying the Mercator projection (\textbf{B}). We show the same projection for a low frequency blip glitch in plots \textbf{C} and \textbf{D}. We note the characteristic thinning of the low frequency blip glitch close to the bottom of plot \textbf{D}. One technical note is that the Mercator transform does alter the frequency labels that are present in the qscans above, however, they can be ignored as Gravity Spy does not read labels.}
\label{fig:mercator}
\end{figure*}

We apply a Mercator projection to the qscans that stretches images vertically~\cite{Pearson1990MapPT}; an example of a simulated GW source with a total mass of $270~M_\odot$ and low frequency blip before and after applying the projection can be seen in Figure \ref{fig:mercator}. 
This projection gives us better resolution of the data above $\sim100$ Hz. 
As shown in Figure \ref{fig:mercator}, the GW (\textbf{A}) presents as a long thick line whereas the low frequency blip (\textbf{D}) tapers off and becomes quite thin. Through the lens of the Mercator projection, these characteristics are common between low frequency blips and GW sources in the $[250M_\odot,350M_\odot]$ range.

We retrained another classifier using Gravity Spy's architecture on a training set of simulated signals in the total mass range $[250M_\odot,350M_\odot]$ and low frequency blip examples with the Mercator transformation applied to the qscans. We call this binary classification model GSpySVG\_EHM. On the same test set that the original Gravity Spy model classified $0/50$ extremely high mass simulated signals correctly, GSpySVG\_EHM was able to correctly classify $48/50$. Moreover, our binary model saw a $100\%$ correct classification rate on $50$ test low frequency blip glitches. GSpySVG\_EHM acts as a proof of concept that building different feature sets with qscan processing techniques can aid in challenging classifications. A first implementation based on the recommendations from this proof-of-principle study leveraged this key result by funneling a candidate’s classification to a low-mass or high-mass classifier based on its initial estimated mass \cite{alvarezlopez2023gspynettree}.

\section{Summary and conclusion}\label{sec_5}

In this paper, we have discussed the types of GW detector noise artifacts that have been known to mimic real GW events and have outlined our investigation into prototyping a reliable signal-vs-glitch classification model using Gravity Spy's architecture.

The main finding of this paper is that breaking up the original Gravity Spy model into multiple classifiers that specialize in differentiating similar CBC signals and detector glitches results in significant improvements for signal-vs-glitch classification accuracy.  
We have prototyped a restructuring of the original Gravity Spy model in the GSpySVG\_LM and GSpySVG\_HM models which have shown great improvements when tested on simulated signals, glitches, and O3b candidate events. 

We compared the performance of our method to using the original Gravity Spy model as a signal-vs-glitch classifier, outside of its designed scope of use. 
In their target mass ranges, GSpySVG\_LM and GSpySVG\_HM achieved an overall simulated signal classification accuracy of $97\%$, an increase from $52\%$ with the original Gravity Spy model. 
Furthermore, GSpySVG\_LM and GSpySVG\_HM classified test sets consisting of 200 glitch examples that often mimic the form of signals with $99\%$ accuracy each. Both GSpySVG\_LM and GSpySVG\_HM were also tested on O3b data where they were able to correctly classify all confirmed events as non-glitches, an improvement from $6/21$ with the original Gravity Spy model. In addition to our main improvements in the $<250M_\odot$ range, we also investigated signals and glitches that may appear with a slight offset in time. For signals with a time offset in the range of $\pm0.1$ seconds, we were able to increase classification accuracy from $83\%$ to $99\%$ by adding random time offsets to each of the training images in the GSpySVG\_LM and GSpySVG\_HM training sets. 

We further improved the performance of our prototype method to high mass CBC sources that share morphology with low frequency blip glitches. We developed novel qscan scaling techniques and demonstrated that Gravity Spy's CNN is able to distinguish between signals with total mass greater than $250M_\odot$ and low frequency blip glitches when a Mercator projection is applied to the qscans. We trained a third binary classifier that made use of the Mercator projection to emphasize distinguishing features between the extremely high mass GW and low frequency blip classes. This GSpySVG\_EHM model classified simulated signals with a total mass between $250M_\odot$ and $350M_\odot$ with $96\%$ accuracy. This is a substantial improvement over $0\%$ accuracy that we saw with the original Gravity Spy model. This part of the project highlights the potential for qscan processing techniques to be leveraged for CNN classifications. 

We note that, at the time of writing, a complete workflow based on this work that combines the GSpySVG\_LM, GSpySVG\_HM, and GSpySVG\_EHM models into a reliable signal-vs-glitch classifier for future GW observing runs has been implemented by Alvarez et al.~\cite{alvarezlopez2023gspynettree}. 

\textit{Acknowledgements:} SJ was supported by the NSERC USRA program. SS acknowledges support from the United States National Science Foundation (NSF) under award PHY-1764464 to the LIGO Laboratory and NSF grant PHY-1806656. JM was supported by the Canada Research Chairs programs. DD is supported by the NSF as a part of the LIGO Laboratory. LIGO was constructed by the California Institute of Technology and Massachusetts Institute of Technology with funding from the National Science Foundation, and operates under cooperative agreement PHY-1764464. This material is based upon work supported by NSF's LIGO Laboratory which is a major facility fully funded by the National Science Foundation. The authors are grateful for computational resources provided by the LIGO Laboratory and supported by National Science Foundation Grants PHY-0757058 and PHY-0823459.

% \section{References}

\bibliography{bib}

\end{document}